\documentclass[12pt]{article}
\usepackage{graphicx}
\usepackage{amsmath}
\usepackage{amssymb}
\usepackage{amscd}

\usepackage{theorem}
\theoremstyle{plain}
\newtheorem{thm}{Theorem}[section]
\newtheorem{prp}[thm]{Proposition}
\newtheorem{dfn}[thm]{Definition}
\newtheorem{cor}[thm]{Corollary}

\makeatletter
 
 \@addtoreset{equation}{section}
\makeatother

\newcommand{\Rmath}{{\mathbb{R}}}

\newcommand{\Sets}{{\bf Sets}}
\newcommand{\Ocal}{{\cal O}}

\newcommand{\Obj}{{\rm Obj}}
\newcommand{\Mor}{{\rm Mor}}

\newcommand{\Hhat}{\hat{H}}
\newcommand{\Rhat}{\hat{R}}

\newcommand{\Mbold}{\mathbf{M}}
\newcommand{\Nbold}{\mathbf{N}}
\newcommand{\Kbold}{\mathbf{K}}

\newcommand{\Lbold}{\mathbf{L}}

\newcommand{\Sbold}{\mathbf{S}}
\newcommand{\Tbold}{\mathbf{T}}
\newcommand{\Omegabold}{\mathbf{\Omega}}

\newcommand{\Hom}{{\rm Hom}}
\newcommand{\Sub}{{\rm Sub}}
\newcommand{\dom}{{\rm dom}}
\newcommand{\cod}{{\rm cod}}

\newcommand{\Dcal}{{\mathfrak D}}
\newcommand{\Lcal}{{\mathfrak L}}
\newcommand{\Scal}{\mathcal{S}}
\newcommand{\Hcal}{\mathcal{H}}
\newcommand{\Ccal}{\mathcal{C}}
\newcommand{\Tcal}{\mathcal{T}}
\newcommand{\Vcal}{\mathcal{V}}
\newcommand{\Vmath}{{\mathfrak V}}
\newcommand{\CcalR}{{\mathcal{C}_R}}
\newcommand{\CcalRe}{{\mathcal{C}_R^{e \downarrow}}}
\newcommand{\Ocaltilde}{\mathcal{O}/_{\sim}}
\newcommand{\Fhat}{\hat{F}}
\newcommand{\Ihat}{\hat{I}}

\newcommand{\pihat}{\hat {\pi}}

\newcommand{\Ghat}{\hat{G}}

\newcommand{\Com}{{\rm Com}}
\newcommand{\ES}{{\rm ES}}

\newcommand{\Abold}{{\bf A}}

\newcommand{\Bcal}{\mathcal{B}}

\newcommand{\vertin}{\mathop{{\rotatebox{90}{$\in$}}}}
\newcommand{\qed}{\hbox{\rule[-2pt]{6pt}{6pt}}}
\begin{document}

\title{Topos-Theoretic Extension of a Modal Interpretation of Quantum Mechanics}

\author{Kunji Nakayama\footnote{email: nakayama@law.ryukoku.ac.jp} \\
\\
Faculty of Law\\
Ryukoku University \\
Fushimi-ku\\
Kyoto 612-8577}


\maketitle

\begin{abstract}

This paper deals with topos-theoretic truth-value valuations
of quantum propositions.
Concretely, a mathematical framework of a specific type of modal approach is extended to
the topos theory,
and further, structures of the obtained truth-value valuations are investigated. 
What is taken up is the modal approach based on a determinate lattice $\Dcal(e,R)$,
which is a sublattice of the lattice $\Lcal$ of all quantum propositions
and is determined by a quantum state $e$ and a preferred determinate observable $R$. 
Topos-theoretic extension is made in the functor category $\Sets^{\CcalR}$ 
of which base category $\CcalR$ is determined by $R$.
Each true atom, which determines truth values, true or false,
of all propositions in $\Dcal(e,R)$, generates also a multi-valued valuation function 
of which domain and range are $\Lcal$ and a Heyting algebra 
given by the subobject classifier in $\Sets^{\CcalR}$,
respectively.
All true propositions in $\Dcal(e,R)$ are assigned the top element of the 
Heyting algebra by the valuation function.
False propositions including the null proposition
are, however, assigned values larger than the bottom element. 
This defect can be removed by use of a subobject semi-classifier.
Furthermore, in order to treat
all possible determinate observables in a unified framework,
another valuations are constructed in the functor category  $\Sets^{\Ccal}$.
Here, the base category $\Ccal$ includes 
all $\CcalR$'s as subcategories.
Although $\Sets^{\Ccal}$ has a structure apparently 
different from $\Sets^{\CcalR}$,
a subobject semi-classifier of $\Sets^{\Ccal}$ gives 
valuations completely equivalent to
those in $\Sets^{\CcalR}$'s.
\end{abstract}

\newpage

\section{Introduction}
\label{intro}
Although quantum mechanics has achieved marvelous success,
its foundations or interpretations are still debatable.
The standard instrumentalism
with emphasis on measurements by an observer 
external to a quantum system as an object
is inappropriate at least for quantum cosmologies
which deal with the universe as a quantum system.
From this viewpoint,
a realism formulation or interpretation 
where `observables' are treated as `beables'
which possess values is desirable,
because such a formulation or interpretation
does not need external observers.

As is well-known, however,
a simple realist's view that
at each state any physical quantity has a value,
or equivalently,
any quantum proposition stating that
an observable has a value in a Borel subset of $\Rmath$
has a determinate truth-value, true or false,
is prohibited by Kochen-Specker's theorem~\cite{KS67}.
In the so-called modal interpretations
which accept realism of physical quantities, therefore, 
only a part of quantum propositions
are given truth-values to avoid Kochen-Specker contradiction. 
(For detailed descriptions of modal interpretations,
see, e.g.,~\cite{vanFraassen91,Bub97,Vermaas99}
and references therein).

On the other hand,
Isham and his collaborators~\cite{Isham97,IB98,BI99,HBI00,BI02,Isham07} 
explored `neo-realism' formulation based on toposes
which are categories satisfying particular properties.
In particular, in a series of papers~\cite{IB98,BI99,HBI00,BI02},
they gave a topos-theoretic representation 
of the Kochen-Specker's theorem
and found out an alternative way to assign a truth value to
any quantum proposition without the contradiction.
Each topos has Heyting-algebra structures built-in,
which are explicitly reflected by a particular object
called a subobject classifier (e.g.,~\cite{Goldblatt79,MM92}).
Utilizing this structure, 
they constructed truth-value functions
defined on all quantum propositions.
The valuations are, therefore, not 2-valued
but multi-valued allowing partly true propositions between 
the false and the true.

Application of the topos theory is
further developed in a series of papers~\cite{DI07a,DI07b,DI07c,DI07d} 
by D\"{o}ring and Isham.
The representation power of categorical logics (e.g.,~\cite{LS86,JLBell88})
enables them to use the topos theory as a new basic language
alternative to the set theory
for mathematical theories of physics.
D\"{o}ring and Isham's project might actually liberate quantum mechanics 
from the observer-object dichotomy,
an origin of notorious paradoxes of quantum mechanics,
and provide proper foundations of quantum cosmology and quantum gravity.

Our primary interest is in relation between 
the modal interpretations and 
the neo-realism.
If the topos theory can be proper framework of quantum mechanics,
it can be expected that appropriately formulated modal interpretations
are actually not interpretations but parts or defectives of the topos-theoretic
quantum mechanics.
Motivated by the interest,
we address two subjects in the present paper.
One is a topos-theoretic extension of a specific type of
modal formulation.
We take up a modal formulism by  Bub~\cite{Bub92}
and, by use of mathematical ingredients therein,
we construct two kinds of topos-theoretic
valuation functions.
The other is investigation of structures 
of the topos-theoretic valuations;
in particular, we make a detailed analysis 
of structural relations between the two valuation functions.

This paper is organized as follows.
In the next section
we briefly review Bub's construction of 
2-valued valuation functions defined on 
a sublattice of quantum propositions,
$\Dcal(e,R)$, which is uniquely determined by
a quantum state $e$ and a determinate observable $R$.
In Section \ref{sec:Topos-Theoretic Valuations},
a topos-theoretic extension of Bub's construction is given.
We define a base category $\CcalR$ 
which is determined by $R$.
Truth-value valuations are constructed in 
the functor category $\Sets^{\CcalR}$.
They are given by characteristic morphisms 
corresponding to `true' subobjects of 
the object $\Lbold$ representing the lattice of all quantum propositions.
The true subobjects are defined by the analogous way 
that the true propositions in the Bub's modal interpretation are determined.
Any quantum proposition takes its truth-value
on the subobject classifier which is a Heyting algebra.
It is, however, shown that 
the valuation functions do not satisfy 
the null-proposition condition proposed by Isham and Butterfield~\cite{IB98}.
We therefore introduce a notion of subobject semi-classifiers
to make the valuation functions
fulfill the null-proposition condition.
In Section \ref{sec:Valuations in Extended}, we work on a base category
$\Ccal$ which includes all $\CcalR$ as subcategories.
Truth-value valuations are constructed
in the functor category $\Sets^{\Ccal}$.
Section \ref{sec:Relation between Valuation} is devoted to clarify
the relation between valuation structures in
$\Sets^{\CcalR}$ and $\Sets^{\Ccal}$.
We prove that the subobject classifier of $\Sets^{\Ccal}$
has a subobject of which components are Heyting algebras
isomorphic to the subobject classifiers of $\Sets^{\CcalR}$'s.
This subobject is a subobject semi-classifier
on which the valuation functions take truth-values.
Thus, it is shown that
the functor categories $\Sets^{\Ccal}$ and $\Sets^{\CcalR}$'s
give actually equivalent truth-value valuations
of quantum propositions.
Our results are summarized in Section \ref{Conclusion} with comments.


\section{Two-Valued Valuations on Determinate Sublattice}
\label{sec:Two-Valued Valuations}
Bub~\cite{Bub92} introduced a maximal determinate sublattice
of quantum propositions 
on which two-valued valuation can be defined
without generating Kochen-Specker contradiction,
starting from a preferred determinate observable
which is supposed to have one of its eigenvalues
even if the quantum system is 
not in the corresponding eigenstate.
He developed a modal interpretation
based on the formulism 
(\cite{Bub97} and references therein).
Below we describe its construction
for the later convenience of reference.

We deal with cases where physical systems
are described in the $n$-dimensional Hilbert Space $\Hcal$.
The set of all rays in $\Hcal$
corresponding to quantum states is denoted by $\Scal$,
and the set of all observables by $\Ocal$.
We choose arbitrarily 
a preferred determinate observable $R \in \Ocal$.
The corresponding self-adjoint operator
on $\Hcal$ is denoted by $\Rhat$.

Let us define the set of all eigenspaces $r_{1}$, $\cdots$, $r_{m}$ 
($m \leq n$) of $\Rhat$ by $\ES(R)$.
For any ray $e \in \Scal$,
a projection of $e$ on the eigenspace $r$
is denoted by $e_{r}$:
\begin{equation}
e_{r} := ( e \vee r^{\bot } ) \wedge  r.
\end{equation} 
If $e_{r}$ is not the zero-space $\{0\}$,
it is a ray. 
The set of such non-zero $e_{r}$'s is denoted by $A(e)$:
\begin{equation}
A(e):= 
\{e_{r}: e_{r} \neq \{0\},\; r \in \ES(R)\}
=
\{ e_{r_1},\cdots,e_{r_k}\},
\end{equation}
where $k \leq m$.
%

The lattice of all quantum propositions,
i.e., the lattice of all subspaces of $\Hcal$,
is denoted by $\Lcal$.
The determinate maximal sublattice $\Dcal(e,R)$ of $\Lcal$ 
is constructed by means of $A(e)$;
that is, $\Dcal(e,R)$ is a lattice $\Lcal_{A(e)}$ 
generated by $k$ orthogonal rays $e_{r_i} $
and all the rays in the subspace 
$\left(\bigvee A(e) \right)^{\bot } $
orthogonal to the $k$-dimensional subspace 
spanned by the elements of $A(e)$. 
Therefore, it is characterized as the commutant in $ \Lcal$ 
of $ \{ e_{r_1}, \cdots, e_{r_k} \} $,
that is,
\begin{equation} 
\Dcal(e,R) = \{ P \in \Lcal : e_{r} \le P \; {\rm or} 
\; e_{r} \le P^{\bot },\; e_{r} \in A(e) \}.
\end{equation}

If the physical system is in $e$ and the observable $R$ has
an eigenvalue corresponding to the eigenspace $r$, 
the associated atom $e_{r} \in A(e)$ should be regarded as true,
and hence,
other atoms of $\Dcal(e,R)$ except for $e_{r}$ 
should be false.
Therefore, all observables of which eigenspaces are spanned by
rays in $\Dcal(e,R)$ are determinate;
if $e_{r}$ is a true atom,
any such observable possesses its eigenvalue corresponding to
the eigenspace including $e_{r}$. 
In addition, the `true' or `false' assignment to the atoms 
defines a truth-value valuation on $\Dcal(e,R)$ via order relations.
That is, for each $e_{r} \in A(e)$ chosen as a true atom,
the two-valued lattice-homomorphism,
$ V^{e_{r}}: \Dcal (e,R) \to \{0,\;1\}$, 
is defined by,
\begin{equation}
V^{e_{r}} (P) :=
\begin{cases}
1 & (P \ge e_{r})  \\
0 & ({\rm otherwise})  \\
\end{cases}
.
\end{equation}
Topos-theoretic extension of this construction is our subject.


\section{Topos-Theoretic Valuations Equipped with a Preferred Determinate Observable}
\label{sec:Topos-Theoretic Valuations}

\subsection{Base Category $\CcalR$ and Functor Category $\Sets^{\CcalR}$}
\label{subsec:Base Category}
In this section, 
we construct valuation functions
defined on the lattice $\Lcal$ of all quantum propositions
for a given preferred determinate observable $R$
and each true atom $e_{r}$.

At the beginning, we describe our rough idea.

Let us consider a quantum proposition $P \in \Lcal$.
If $P \geq e_{r}$, we regard $P$ as true,
whether it belongs to $\Dcal(e,R)$ or not.
On the other hand, even if $P \not\geq e_{r}$,
we do not think of $P$ as false
provided $\pihat_{r}(P)$ is not the zero space $\{0\}$;
it is regarded as partly true.
Here, $\pihat_{r}$ is the projection operator to 
the eigenspace $r \in \ES(R)$.
Since we are provided the determinate observable $R$,
degree of truth of such $P$ should be quantified 
by means of ingredients related to $R$. 
So, we utilize the set $\Com(R)$ 
which is a commutant of the self-adjoint operator $\Rhat$
corresponding to the observable $R$.
That is, we transform $P$ and $e_{r}$ by each $\Fhat \in \Com(R)$,
and define a set $\Vcal(P)$ by
\begin{equation}
\Vcal(P) := \{ \Fhat \in \Com(R):\; \Fhat(P) \geq \Fhat(e_{r}) \}.
\end{equation}
We would like to regard $\Vcal(P)$ as a truth value of 
the proposition $P$.
In fact, $\Vcal$ has desirable properties as a truth-value valuation.
For example, for any $P_{1}$, $P_{2} \in \Lcal$, 
we have
\begin{equation}
P_{1} \leq P_{2} \;\Longrightarrow\; \Vcal(P_{1}) \subseteq \Vcal(P_{2}).
\end{equation}
Further, we have
\begin{equation}
P \geq e_{r} \;\Longrightarrow\; \Vcal(P) = \Com(R),
\end{equation}
and
\begin{equation}
P \not\geq e_{r},\; \pihat_{r}(P) \neq \{0\} 
\;\Longrightarrow\; 
\Vcal(\{0\}) \subset  \Vcal(P) \subset \Com(R).
\end{equation}
These properties suggest that $\Vcal$
gives a multi-valued truth-value valuation
of which target is a logical space 
preordered by the inclusion relation of sets.

In order to realize the above-mentioned idea
in a canonical way,
we utilize topos structure of the functor category $\Sets^{\CcalR}$.
Here, objects of the base category $\CcalR$ are rays;
that is,
\begin{equation}
\Obj(\CcalR) := \Scal.
\end{equation}
The collection of morphisms, $\Mor(\CcalR)$,
is given by the disjoint union of all hom-sets,
$\Hom_{\CcalR} (e,e')$,
which are defined by,
\begin{equation}
\Hom_{\CcalR} (e,e') 
:=
\{\Fhat \in \Com(R): e' = \Fhat e\},
\end{equation}
for each $e$, $e' \in \Obj(\CcalR)$.
The identity morphisms are defined by $e \xrightarrow{1_{e}} e := \Ihat$ 
for any $e$.
The composition of morphisms 
$ e \xrightarrow{\Fhat} e'$ and $ e' \xrightarrow{\Fhat'} e''$ is defined
by $\Fhat' \circ \Fhat := \Fhat' \Fhat \in \Hom_{\CcalR} (e,e'') $.
It is clear that this definition of composition
satisfies the associativity axiom of categories.

As is seen in the last section,
the key ingredient to construct
the 2-valued valuation functions $V^{e_{r}}$
is the subset $A(e)$ of $\Dcal(e,R)$.
Motivated by this, 
we construct topos-theoretic counterparts 
of $A(e)$ and the lattice $\Lcal$
in the functor category $\Sets^{\CcalR}$.
To do so,
we note that 
any $r \in \ES(R)$ is invariant 
under the action of any $\Fhat \in \Com(R)$;
i.e., $\Fhat(r) \subseteq  r$. 
From this property, we can show that
\begin{equation}
\Fhat(e_{r}) = (\Fhat e)_{r}, \label{eq:Fhater}
\end{equation}
for any $e_{r} \in A(e)$.
Here, we allow both sides of equation (\ref{eq:Fhater})
to be the zero-space $ \{ 0 \} $.

Equation (\ref{eq:Fhater}) implies that
we can extend the function $A: \Obj(\CcalR) \to \Obj(\Sets)$ 
to a functor from $\CcalR$ to $\Sets$
by augmenting each $A(e)$ with the zero-space $\{ 0 \}$
as follows:
\begin{equation}
e 
\; \longmapsto \;
\Abold(e) 
:=
A(e)\cup \{\{0\}\},
\end{equation}
\begin{equation}
e \xrightarrow{\displaystyle \Fhat} e'
\; \longmapsto \;
\begin{array}{ccc} \Abold(e)  & \xrightarrow{\displaystyle \Abold( \Fhat)} & \Abold(e')  \\
[-1pt] \vertin &                 & \vertin \\
[-1pt] e_{r}  & \longmapsto   & \Fhat (e_{r}  )
\end{array}
\;.
\end{equation}

As a counterpart of $\Lcal$,
we define a functor $\Lbold:\CcalR \to \Sets$ 
which gives $\Lcal$ for each $e$
and an order homomorphism for each $e \xrightarrow{\Fhat} e'$:
\begin{equation}
e 
\; \longmapsto \; 
\Lbold(e) 
:=
\Lcal,
\end{equation}
\begin{equation}
e \xrightarrow{\displaystyle \Fhat} e'
\; \longmapsto \;
\begin{array}{ccc} \Lbold(e)  & \xrightarrow{\displaystyle  \Lbold(\Fhat)} & \Lbold(e')  \\
[-1pt] \vertin &                 & \vertin \\
[-1pt] P  & \longmapsto   & \Fhat P
\end{array}
\;.
\end{equation}
Note that $\Lbold$ includes $\Abold$ as a subobject;
that is, $\Abold(e) \subseteq \Lbold (e)$
and for any $\Fhat \in \Hom_{\CcalR}(e,e')$,
$\Abold (\Fhat) = \Lbold (\Fhat) |_{\Abold (e)}$.

As a topos, the functor category $\Sets^{\CcalR}$ 
has a particular functor,
the subobject classifier $\Omegabold$.
It is defined by
\begin{equation}
e 
\; \longmapsto \; 
\Omegabold(e) := \{ S: S \mbox{ is a sieve on } e \},
\end{equation}
\begin{equation}
e \xrightarrow{\displaystyle \Fhat} e'
\; \longmapsto \;
\begin{array}{ccc} \Omegabold(e)  & \xrightarrow{\displaystyle  \Omegabold(\Fhat)} & \Omegabold(e')  \\
[-1pt] \vertin &                 & \vertin \\
[-1pt] S  & \longmapsto   & \left\{ \Fhat' \in \Mor(\CcalR) : \Fhat' \circ \Fhat  \in S \right\}
\end{array}
\;. \label{eq:OmegaboldFhat}
\end{equation}
Here, a sieve on $e$ is a set $S$ of such morphisms of which domains are $e$
that if $ \Fhat \in S $, $ \Fhat' \in \Mor(\CcalR)$
and $\Fhat' \circ \Fhat$ exists, then $\Fhat' \circ \Fhat \in S $.
For each $e \in \Obj(\CcalR)$, 
$\Omegabold (e)$ possesses a Heyting-algebra structure
defined by the inclusion relation among sieves,
with the top element,
\begin{equation}
\top_{e} := \bigcup\limits_{e' \in \Scal} \Hom_{\CcalR}(e,e'),
\label{eq:top}
\end{equation}
and the bottom element,
\begin{equation}
\bot_{e} := \emptyset .
\label{eq:bottom}
\end{equation}
It is easy to see that the set function
$\Omegabold(\Fhat):\Omegabold(e) \to \Omegabold(e')$
given by (\ref{eq:OmegaboldFhat}) maps any sieve in $\Omegabold(e)$ 
to a sieve in $\Omegabold(e')$.
Therefore, $\Omegabold$ is well-defined as a functor from $\CcalR$ to $\Sets$.

The subobject classifier $\Omegabold$ is 
a generalization of ${\bold 2}:=\{0,1\} \in \Obj(\Sets)$;
as any subset of a set is determined by
a characteristic function from the set to ${\bold 2}$,
any subobject of an object in $\Sets^{\CcalR}$ 
is determined by a characteristic morphism (i.e., a natural transformation)
from the object to $\Omegabold$. 
That is, 
for any subobject $\Sbold$ of the object $\Lbold$,
(i.e., for any monomorphism $m \in \Hom_{\Sets^{\CcalR}}(\Sbold,\Lbold)$),
there exists one and only one natural transformation
$\chi^{m} \in \Hom_{\Sets^{\CcalR}}(\Lbold,\Omegabold)$
making the following diagram a pullback
in the functor category $\Sets^{\CcalR}$: 
\begin{equation}
\begin{CD}
\Sbold @ > ! >> {\bf 1}\\
   @ V m VV @ VV \tau V \\
\Lbold  @ >> \chi^{m} > \Omegabold
\end{CD}
\; \; .
\label{eq:Sboldpullback}
\end{equation}
Conversely, any $\chi \in \Hom_{\Sets^{\CcalR}}(\Lbold,\Omegabold)$
determines, up to isomorphism, 
monomorphisms $m$ with $\cod(m) = \Lbold$
making the diagram (\ref{eq:Sboldpullback}) a pullback.
(We hereafter deal with $m$ as the inclusion morphism 
$\iota^{\Sbold \Lbold}$ from $\Sbold$ into $\Lbold$; 
namely, a natural transformation which yields 
a set-theoretic inclusion function 
$\iota^{\Sbold \Lbold}_{e} : \Sbold(e) \hookrightarrow  \Lbold(e)$
for each $e$.
Correspondingly, $\chi^{m}$ is written as $\chi^{\Sbold \Lbold}$.)
In the above diagram, 
the functor ${\bf 1}$ is a final object of $\Sets^{\CcalR}$,
which assigns each $ e \in  \Obj(\CcalR)$ and
each $ e \xrightarrow{ \Fhat } e' \in \Mor (\CcalR)$ 
the one-point set $ {\bold 1} (e) := \{ \ast  \}$ 
and the identity ${\bold 1} (e \xrightarrow{\Fhat} e') := {\rm id}_{\{\ast \}}$,
respectively.
The morphism $ \tau $,
which is often called a {\it true}, is a global element of $\Omegabold$
taking the top element of $ \Omegabold(e) $ for each $e$; 
${\tau}_{e}(*) := \top_{e}$. 
Here, in general,
a global element of $\Kbold \in \Obj(\Sets^{\CcalR})$ is defined as
a natural transformation $\mu : {\bf 1} \xrightarrow{\centerdot} \Kbold$;
that is, for each $e$, it chooses one element $\mu_{e}(*) \in \Kbold(e)$
in such a way that the naturality diagram,
\begin{equation}
\begin{CD}
{\bf 1} (e)  @ > {\mu}_{ e} >> \Kbold (e) \\
   @ V {\bf 1} (\Fhat)  VV  @ VV \Kbold (\Fhat) V \\
{\bf 1} (e') @ > {\mu}_{ e'} >> \Kbold (e')   
\end{CD} 
\;  \quad,
\label{eq:globalelement}
\end{equation}
commutes for any $\Fhat \in \Hom(e,e')$.
Finally, the morphism 
$\chi^{\Sbold \Lbold}:\Lbold \xrightarrow{\centerdot} \Omegabold $ 
is a natural transformation defined by
\begin{equation}
\chi^{\Sbold \Lbold}_{e}(P)
:=
\{ \Fhat \in \Mor(\CcalR) : 
\dom(\Fhat) = e, \; \Lbold (\Fhat) (P) \in \Sbold (\cod (\Fhat)) \}
\in \Omegabold(e) ,
\end{equation}
for any $P \in \Lbold(e) = \Lcal$.

For each object $e$,
the diagram (\ref{eq:Sboldpullback}) in the topos $\Sets^{\CcalR}$ 
reduces to a pullback diagram in $\Sets$:
\begin{equation}
\begin{CD}
\Sbold(e) @ > ! >> {\bf 1}(e)\\
   @ V \iota^{\Sbold \Lbold}_{e} VV @ VV \tau_{e} V \\
\Lbold(e)  @ >> \chi^{\Sbold \Lbold}_{e} > \Omegabold(e)
\end{CD}
\; \; .
\end{equation}
The sieve $\chi^{\Sbold \Lbold}_{e}(P) \in \Omegabold(e)$
indicates nearness of the proposition $P$ at the stage $e$
to the subobject $\Sbold$.
In fact, if $P \in \Sbold(e)$, 
then $\Lbold (\Fhat) (P) \in \Sbold (\cod (\Fhat)) $ 
for all $\Fhat$ of which domain is $e$.
Therefore, $\chi^{\Sbold \Lbold}_{e} (P)$ equals 
the top element $\top_{e}$ of $\Omegabold(e)$.
On the other hand, 
if $P \notin \Sbold(e')$ for any $e'$ satisfying 
$\Hom_{\CcalR}(e,e') \neq \emptyset $,
then $\chi^{\Sbold \Lbold}_{e} (P) = \emptyset$,
the bottom element $\bot_{e}$ of $\Omegabold(e)$.
If $P \notin \Sbold(e)$ but there exists some $\Fhat \in \Mor(\CcalR)$
such that $\dom(\Fhat) = e$ and $\Lbold (\Fhat) (P) \in \Sbold (\cod (\Fhat)) $,
then $\chi^{\Sbold \Lbold}_{e} (P)$ is a sieve on $e$
between $\bot_{e}$ and $\top_{e}$.
Furthermore, for $P$, $Q \in \Lbold(e)$,
if $\Lbold (\Fhat) (P) \in \Sbold (\cod (\Fhat)) $ implies 
$\Lbold (\Fhat) (Q) \in \Sbold (\cod (\Fhat)) $ for any $\Fhat$,
which intuitively means that $Q$ is closer to the subobject $\Sbold$
than $P$,
then $\chi^{\Sbold \Lbold}_{e} (P) \leq \chi^{\Sbold \Lbold}_{e} (Q)$.
In the sense that, the closer the proposition in is to $\Sbold$,
the larger the assigned sieve becomes,
$\chi^{\Sbold \Lbold}_{e}$ acts as an indicator of 
nearness of any proposition in $\Lbold(e)$
to $\Sbold$ at the stage $e$.

\subsection{Prerequisites for True Subobjects and Valuation Functions}
\label{subsec:Prerequisites for True}
If a subobject $\Sbold$ of $\Lbold$ consists of true propositions,
its characteristic morphism $\chi^{\Sbold \Lbold}_{e}$
indicates how close a proposition $P \in \Lbold(e)=\Lcal$ is 
to the true propositions at the stage $e$.
It defines, therefore, a generalized truth-value valuation functions. 

In order for $\Sbold$ to represent truth,
or equivalently, for $\chi^{\Sbold \Lbold}_{e}$ 
to be a truth-value valuation function,
each set $\Sbold(e)$ of propositions should be a filter in $\Lbold(e)$.
That is, for each $e \in \Obj(\CcalR)$,
implication relations
\begin{equation}
P \in \Sbold(e),
\;
Q \in \Lbold(e),
\;
P \leq Q
\;  
\Longrightarrow
\;
Q \in \Sbold(e)
\label{eq:filter1}
\end{equation}
and
\begin{equation}
P, \; Q \in \Sbold(e)
\;
\Longrightarrow
\;
P \wedge Q \in \Sbold(e)
\label{eq:filter2}
\end{equation}
should be satisfied.
They are abstraction from the characteristic that
collections of all true propositions should satisfy.
(For detailed properties of filters, 
see, e..g., Davey and Priestly~\cite{DP90}.)

Implication relation (\ref{eq:filter1}), which means that $\Sbold(e)$ is an up-set,
implies that if a proposition $P$ is true and $P$ always implies $Q$,
then $Q$ must be true.
If $\Sbold(e)$ is an up-set for each $e \in \Obj(\CcalR)$, 
then, for any $e \in \Obj(\CcalR)$ and any $P$, $Q \in \Lbold(e)=\Lcal$, 
the characteristic map
$\chi^{\Sbold \Lbold}_{e}$ satisfies the monotonicity condition
proposed by Isham and Butterfield~\cite{IB98},
\begin{equation}
P \leq Q
\;
\Longrightarrow
\;
\chi^{\Sbold \Lbold}_{e}(P)
\leq 
\chi^{\Sbold \Lbold}_{e}(Q),
\end{equation}
which any valuation function must satisfy.

The second condition (\ref{eq:filter2}),
which means that
$\Sbold(e)$ is closed under the meet ($\wedge $) operation,
implies that, if propositions $P$ and $Q$ are true, 
so is their conjunction.
From this condition,
we can derive the exclusivity condition
proposed by Isham and Butterfield~\cite{IB98}.
Note that, from the relation (\ref{eq:filter2}),
we have
\begin{equation}
P \wedge Q \notin \Sbold(e),
\;
P \in \Sbold(e)
\;
\Longrightarrow
\;
Q \notin \Sbold(e) 
.
\end{equation} 
Therefore, the exclusivity condition is proved as
\begin{eqnarray}
\chi^{\Sbold \Lbold}_{e} (P \wedge Q) <  \top_{e},
\;
\chi^{\Sbold \Lbold}_{e} (P) =  \top_{e} 
&
\Longrightarrow
&
P \wedge Q \notin \Sbold(e),
\;
P \in \Sbold(e)
\nonumber\\
\;
&
\Longrightarrow
&
Q \notin \Sbold(e) 
\nonumber\\
&
\Longrightarrow
&
\chi^{\Sbold \Lbold}_{e} (Q) <  \top_{e} 
.
\end{eqnarray} 

With regard to conditions that generalized valuation functions should satisfy,
Isham and Butterfield~\cite{IB98} proposed the unit-proposition condition,
the null-proposition condition, and the functional composition condition
besides the above-mentioned ones.
Among them, the first two conditions concern our formulism.
We check them after obtaining our valuation functions. 
At the moment, we propose only (\ref{eq:filter1}) and (\ref{eq:filter2}).

\subsection{Construction of True Subobjects and Valuation Functions}
\label{subsec:Construction of True}
As described in Section \ref{sec:Two-Valued Valuations},
true propositions in the lattice $\Dcal(e,R)$
are given by a true atom $e_{r} \in A(e)$.
In the current case, analogously,
the true subobject of $\Lbold$
is determined by a global element 
$\sigma : {\bf 1} \xrightarrow{\centerdot} \Abold $
which specifies true atoms for 
all $e \in \Obj(\CcalR)$ simultaneously.

For any $r \in \ES(R)$,
we define a map 
$\sigma^{r} : \Scal \rightarrow \Scal \cup \{\{0\}\} $ 
by $\sigma^{r}(e)=e_{r}$.
Note that, for each $e \in \Scal$,
$\sigma^{r}_{e} := \sigma^{r}(e)$ can be regarded as a map 
$ \sigma^{r}_{e}:\{*\} \rightarrow \Abold(e)$
defined by
\begin{equation}
\sigma^{r}_{e}(*):=e_{r},
\label{eq:sigmare}
\end{equation}
where $e_{r}$ is allowed to be the zero-space $\{ 0 \}$.
\begin{prp}
For each $r \in \ES(R)$, 
the map $ \sigma^{r} $ is a global element of the functor $\Abold:\CcalR \rightarrow \Sets$;
namely, it is a natural transformation 
$\sigma^{r} : {\bold 1} \xrightarrow{\centerdot} \Abold$
\end{prp}
{\bf Proof.}
From equation (\ref{eq:sigmare}),
it follows that 
\begin{equation}
 \sigma^{r}_{e'} (*) = e'_{r} = (\Fhat e)_{r} 
= \Fhat (e_{r}) = 
\Fhat ({\sigma}^{r}_{ e}(*) )=\Abold(\Fhat) ({\sigma}^{r}_{ e}(*) ),
\end{equation}
for any $e$, $e' \in \Obj(\CcalR)$ and 
$\Fhat \in \Hom_{\CcalR}(e,e')$.
Thus, $\sigma^{r}$ satisfies the commutative diagram (\ref{eq:globalelement})
which defines a global element.
\hfill\qed

For any $r \in \ES(R)$,
we define a subobject $\Tbold^{r} : \Ccal \rightarrow \Sets$
of $\Lbold$ by means of 
the global element $\sigma^{r}: {\bf 1} \xrightarrow{\centerdot} \Abold$
as follows:
\begin{equation}
e
\;
\longmapsto
\;
\Tbold^{r} (e) 
:= 
\{P \in \Lbold(e): P \ge \sigma^{r}_{e} (*)\},
\end{equation}
\begin{equation}
e \xrightarrow{ \displaystyle \Fhat} e'
\; \longmapsto \;
\begin{array}{ccc} \Tbold^{r} (e)  & \xrightarrow{\displaystyle  \Tbold^{r} (\Fhat)} & \Tbold^{r} (e')  \\
[-1pt] \vertin &                 & \vertin \\
[-1pt] P  & \longmapsto   & \Fhat P
\end{array}
\;.
\end{equation}
Here, 
the function $ \Tbold^{r}(\Fhat) :\Tbold^{r}(e) \rightarrow \Tbold^{r} (e')$ is well-defined,
because
\begin{eqnarray}
P \in \Tbold^{r}(e) 
& \Longleftrightarrow &
P \ge \sigma^{r}_{e}(*)  \nonumber\\
& \Longrightarrow &
\Fhat P \ge  \Fhat (\sigma^{r}_{e}(*)) = \sigma^{r}_{e'} (*) \nonumber\\
& \Longleftrightarrow &
\Fhat P \in \Tbold^{r}(e') 
.
\end{eqnarray}
Furthermore,
it is easy to see that $\Tbold^{r} $ is a filter.
Thus, it can be a true subobject of $\Lbold$.

Suppose a physical system is in a state $e$,
and that $e_{r}$ ($\ne \{0\}$) is a true atom
of $\Dcal(e,R)$. 
Then,
we define the corresponding truth-value $ {\Vcal}^{r}_{e}(P) \in \Omegabold(e)$ 
of any $P \in \Lbold(e) = \Lcal$
at the stage $e$ by
\begin{eqnarray}
{\Vcal}^{r}_{e}(P)
& := &
\chi^{\Tbold^{r} \Lbold}_{e}(P) \nonumber\\
& = &
\bigcup\limits_{e' \in \Scal}
\{ \Fhat \in \Hom_{\CcalR}(e,e')
:\Lbold(\Fhat)(P) \in \Tbold^{r}(\Fhat)(e) \} \nonumber\\
& = &
\bigcup\limits_{e' \in \Scal}
\{ \Fhat \in \Hom_{\CcalR}(e,e')
: \Fhat(P) \geq  e'_{r} \} .
\label{eq:valuation2}
\end{eqnarray}

In the following,
we describe some properties that $\Vcal^{r}_{e}$ satisfies.
\begin{prp}
If $P \in \Dcal(e,R)$ and $V^{e_r}(P)=1$,
then, 
\begin{equation}
\Vcal^{r}_{e} (P) = \bigcup\limits_{e' \in \Scal} \Hom_{\CcalR}(e,e') = \top_{e} \in \Omegabold(e).
\end{equation}
\end{prp}
{\bf Proof.}
We have the following implication relations:
\begin{eqnarray}
V^{e_{r}}(P)=1
& \Longrightarrow &
P \geq e_{r} 
\nonumber \\
&  \Longrightarrow &
\forall e' \in \Obj(\CcalR),\;
\forall \Fhat \in \Hom_{\CcalR}(e,e'), \;
\Fhat P \geq \Fhat(e_{r}) = e'_{r}
\nonumber \\
& \iff &
\Vcal^{r}_{e} (P) = \bigcup\limits_{e' \in \Obj(\CcalR)} \Hom_{\CcalR}(e,e') = \top_{e}. 
\end{eqnarray}
\hfill\qed

The converse of Proposition 32 is not true,
since there can exist propositions $P$
such that $\Vcal^{r}_{e}(P)= \top_{e}$ and $P \not\in \Dcal(e,R)$.
On the other hand,
$\Vcal^{r}_{e}$ does not take  
the bottom $\bot_{e} = \emptyset$  of $\Omegabold(e)$;
its minimum value is
\begin{equation}
\bot_{e_{r}} 
:= \bigcup\limits_{e' \in \Scal} \{ \Fhat \in \Hom_{\CcalR}(e,e'): \Fhat (e_{r}) = \{ 0 \} \}
\in \Omegabold(e).
\end{equation}
\begin{prp}
For all $P \in \Lbold(e)$,
\begin{equation}
\Vcal^{r}_{e} (P) 
\geq
\bot_{e_{r}} .
\end{equation}
\end{prp}
{\bf Proof.}
\begin{eqnarray}
e \xrightarrow{\Fhat} e' \in \bot_{e_{r}}
 & \iff & 
\Fhat(e_{r}) = \{0\}
\nonumber \\
 & \Longrightarrow & 
\forall P \in \Lbold(e),\;
\Fhat(P) \geq \Fhat(e_{r})
\nonumber \\
& \iff & 
\forall P \in \Lbold(e),\;
\Fhat \in \Vcal^{r}_{e}(P)
.
\end{eqnarray}
\hfill\qed

Note that, if $P$ belonging to $\Dcal(e,R)$
satisfies $P \subseteq e_{r}^{\bot }$ and $\pihat_{r}(P) \neq \{ 0 \}$,
then $V^{e_{r}}(P)=0$ and $\Vcal^{r}_{e}(P) > \bot_{e_{r}}$;
the valuation function $\Vcal^{r}_{e}$ gives
a finer truth-value assignment to false propositions of $\Dcal(e,R)$,
depending on their nearness to $\Tbold^{r}(e)$.

Let us check the unit-proposition condition 
and the null-proposition condition.
The former is represented 
in terms of our notation as
\begin{equation}
\Vcal^{r}_{e} (I) = \top_{e},
\end{equation}
where $I$ is a unit proposition
which corresponds to the entire Hilbert space $\Hcal$.
It is clear from the definition (\ref{eq:valuation2})
that $\Vcal^{r}_{e}$ satisfies this condition.  
On the other hand, it is not compliant 
to the latter condition.
That is, we have
\begin{equation}
\Vcal^{r}_{e} (\{0\}) = \bot_{e_{r}} > \bot_{e}.
\end{equation}
The essential reason of the noncompliance 
to the null-proposition condition
is the fact that
any ray becomes to $\{0\}$ 
(or equivalently, any state vector vanishes) by action of
some linear operators.
This fact is in common with the normalization issue
occurring in the topos-based interpretation of
state-vector reduction given by Isham~\cite{Isham07}.
To avoid the normalization issue,
Isham proposed mathematical framework
using restricted sets of nonvanishing state vectors.
In order to complete the null-proposition condition,
his approach might be promising also in our case.
In the next subsection, however,
we would like to propose another answer.


\subsection{Valuation Using Subobject Semi-Classifier $\delta_{r} \Omegabold$}
\label{subsec:Valuation Using}
As is seen in the last subsection,
the null-proposition is not assigned the bottom $\bot_{e}$
by the valuation function $\Vcal^{r}_{e}$.
So, we construct alternative targets of
valuation functions in order that
the null-proposition condition is satisfied.

For each $r \in \ES(R)$ and $e \in \Scal$,
we define a set of sieves on $e$, $\delta_{r}\Omega(e)$,
by
\begin{equation}
\delta_{r}\Omega(e) := \{ S \in \Omegabold(e) : \bot _{e_{r}} \subseteq S \}.
\end{equation}
Regarding this,
we have the following proposition:
\begin{prp}
For each $r \in \ES(R)$ and $e \in \Scal$,
the set $\delta_{r} \Omega(e)$ of sieves is 
sublattice of $\Omegabold(e)$.
It is, further, a Heyting algebra.
\end{prp}
{\bf Proof.}
It is easy to see that $\delta_{r} \Omega(e)$
is a sublattice of $\Omegabold(e)$,
i.e.,
it is closed under 
the join ($\vee$) and the meet ($\wedge$) operations.

Also, it is apparent that 
$\delta_{r} \Omega(e)$ has $\top_{e}$ as the top
and $\bot_{e_{r}}$ as the bottom.

We show closure under
the pseudocomplement operation ($\Rightarrow $)
defined in the Heyting algebra $\Omegabold(e)$.
Suppose that $S_{1}$, $S_{2} \in \delta_{r} \Omega(e)$.
Then, $S_{1} \Rightarrow S_{2}$,
which is defined by the maximum sieve $S$ 
such that $S_{1} \wedge S \leq S_{2}$,
is explicitly given by
\begin{eqnarray}
S_{1} \Rightarrow S_{2}
& = &
\left\{
\Fhat \in \Mor(\CcalR): 
\forall \Fhat' \in \Mor(\CcalR) \; 
(\Fhat' \circ \Fhat \in S_{1} \; \Longrightarrow \; \Fhat' \circ \Fhat \in S_{2})
\right\}
\nonumber\\
& \in & 
\Omegabold(e).
\end{eqnarray}
On the other hand, for any $\Fhat' \in \Mor(\CcalR)$,
if $\Fhat \in \bot_{e_{r}}$ and $\Fhat' \circ \Fhat$ is definable,
then $\Fhat' \circ \Fhat (e_{r}) = \{ 0 \}$,
hence, $\Fhat' \circ \Fhat \in \bot_{e_{r}}$.
Therefore, $\bot_{e_{r}} \subseteq S_{1} \Rightarrow S_{2}$,
hence, $S_{1} \Rightarrow S_{2} \in \delta_{r} \Omega(e)$.
Since $S_{1} \Rightarrow S_{2}$ is the maximum $S$ 
such that $S_{1} \wedge S \leq S_{2}$ in $\Omegabold(e)$,
so is it also in $\delta_{r} \Omega(e)$.
Therefore, $S_{1} \Rightarrow S_{2}$ is 
the pseudocomplement in $\delta_{r} \Omega(e)$.
\hfill\qed

Note that, although $\delta \Omega (e)$ itself is a Heyting algebra,
it is not a Heyting subalgebra of $\Omegabold(e)$.
In fact, as was shown, they have different bottom elements.

\begin{prp}
For any $S \in \delta_{r}\Omega(e)$ 
and $\Fhat \in \Mor(\CcalR)$
such that $\dom(\Fhat) = $e,
$\Omegabold(\Fhat)(S) \in \delta_{r}\Omega(e')$.
\end{prp}
{\bf Proof.}
Suppose that $\Fhat' \in \Hom_{\CcalR}(e',e'')$.
Then, it follows that
\begin{eqnarray}
\Fhat' \in \bot_{e'_{r}}
& \Longrightarrow  &
\Fhat' \Fhat (e_{r}) = \Fhat'(e'_{r}) = \{0\}
\nonumber \\
 &  \iff  &
\Fhat' \circ \Fhat \in \bot_{e_{r}}
\nonumber \\
&  \Longrightarrow  &
\Fhat' \circ \Fhat \in S
\nonumber \\
&  \iff  &
\Fhat' \in \Omegabold(\Fhat)(S).
\end{eqnarray}
Thus, $\bot_{e'_{r}} \subseteq \Omegabold(\Fhat)(S)$,
i.e., $\Omegabold(\Fhat)(S) \in \delta_{r}\Omega(e')$.
\hfill\qed

Summarizing Propositions 34 and 35,
we obtain the following theorem:
\begin{thm} For any $r \in \ES(R)$,
the subobject classifier $\Omegabold$ includes 
a subobject $\delta_{r} \Omegabold$ defined by
\begin{equation}
e 
\; \longmapsto \; 
\delta_{r}\Omegabold(e) := 
\delta_{r}\Omega(e),
\end{equation}
and
\begin{equation}
e \xrightarrow{\displaystyle \Fhat} e'
\; \longmapsto \;
\begin{array}{ccc} \delta_{r}\Omegabold(e)  & \xrightarrow{\displaystyle  \delta_{r}\Omegabold(\Fhat)} & \delta_{r}\Omegabold(e')  \\
[-1pt] \vertin &                 & \vertin \\
[-1pt] S  & \longmapsto   & \Omegabold(\Fhat)(S)
\end{array}
\;.
\end{equation}
For each $e$, $\delta_{r}\Omegabold(e)$ is a sublattice of 
$\Omegabold(e)$ and, further, a Heyting algebra
with the top $\top_{e}$ and the bottom $\bot_{e_{r}}$.
\end{thm}

In the following, it is shown that
$\delta \Omegabold$ is a subobject semi-classifier 
defined in Appendix A.
First, note that we can define a natural transformation
$\delta_{r} \tau : 
{\bf 1} \xrightarrow{\centerdot} \delta_{r} \Omegabold$ by 
$\delta_{r} \tau_{e}(*) := \top_{e} 
\in \delta_{r} \Omegabold(e)$.
It is easy to see that the diagram
\begin{equation}
\begin{CD}
{\bf 1} @ = {\bf 1} \\
   @ V {\delta_{r} \tau} VV @ VV  \tau V \\
 \delta_{r} \Omegabold @ >>  \iota^{\delta_{r}\Omegabold \, \Omegabold} >  \Omegabold
\end{CD}
\;\;
\end{equation}
is a pullback.

Next, suppose that $\Mbold \in \Sets^{\CcalR}$
and $\Nbold$ is a subobject of $\Mbold$,
such that,
for any $e \in \Obj(\CcalR)$ and $x \in \Mbold(e)$,
$\bot_{e_{r}} \subseteq \chi^{\Nbold \Mbold}_{e}(x)$,
or equivalently, 
for any $e \in \Obj(\CcalR)$,
\begin{equation}
\chi^{\Nbold \Mbold}_{e}(\Mbold(e)) \subseteq  \delta_{r} \Omegabold(e).
\label{eq:inclusion}
\end{equation}
Then, we have a natural transformation 
$\delta_{r} \chi^{\Nbold \Mbold} : \Mbold \xrightarrow{\centerdot} \delta_{r}\Omegabold$
which is defined by $\delta_{r} \chi^{\Nbold \Mbold}_{e}(x) := \chi^{\Nbold \Mbold}_{e}(x)$
for each $e \in \Obj(\CcalR)$ and $x \in \Mbold(e)$.
Note that $\delta_{r} \chi^{\Nbold \Mbold}$ is
related to $\chi^{\Nbold \Mbold}$ as
\begin{equation}
\chi^{\Nbold \Mbold} 
= 
\iota^{\delta \Omegabold \, \Omegabold} \circ \delta_{r} \chi^{\Nbold \Mbold}.
\end{equation} 
Let $\delta_{r}\Sub(\Mbold)$ be a collection of subobjects 
$\Nbold$ of $\Mbold$ satisfying (\ref{eq:inclusion}).
As a result of the above consideration, 
$\delta_{r} \Omegabold$ is a subobject semi-classifier
of $\delta_{r}\Sub(\Mbold)$.
Thus, we obtain the following theorems which are translations
of Propositions A3 and A4.
\begin{thm}
Suppose that $\Mbold \in \Obj(\Sets^{\CcalR})$ 
and $\Nbold \in \delta_{r}\Sub(\Mbold)$.
Then the diagram
\begin{equation}
\begin{CD}
\Nbold @ >  ! >> {\bf 1} \\
   @ V {\iota^{\Nbold \Mbold}} VV @ VV  \delta_{r} \tau V \\
 \Mbold @ >> \delta_{r} \chi^{\Nbold \Mbold} > \delta_{r} \Omegabold
\end{CD}
\end{equation}
is a pullback.
\end{thm}
\begin{thm}
(i) Suppose that $\Mbold \in \Obj(\Sets^{\CcalR})$ 
and $\Nbold \in \delta_{r}\Sub(\Mbold)$.
If there exists a morphism 
$\Mbold \xrightarrow{\zeta} \delta_{r} \Omegabold$
which makes the diagram
\begin{equation}
\begin{CD}
\Nbold @ >  ! >> {\bf 1} \\
   @ V {\iota^{\Nbold \Mbold}} VV @ VV \delta_{r} \tau V \\
 \Mbold @ >>  \zeta> \delta_{r} \Omegabold
\end{CD}
\;\;
\end{equation}
a pullback,
then $\zeta = \delta_{r} \chi^{\Nbold \Mbold}$.

(ii) Conversely, for each morphism
$\Mbold \xrightarrow{\zeta} \delta_{r} \Omegabold$,
there exists $\Nbold \in \delta_{r}\Sub(\Mbold)$
satisfying $\zeta = \delta_{r} \chi^{\Nbold \Mbold}$, 
up to isomorphism,
hence, making the corresponding diagram (3.41) a pullback.
\end{thm}

These theorems imply that
we can regard
$\delta_{r} \Omegabold$ and $\delta_{r} \chi^{\Nbold \Mbold}:
\Mbold \xrightarrow{\centerdot} \delta_{r} \Omegabold$
as a subobject classifier and a characteristic morphism,
respectively, 
when we consider 
only a class of subobjects included in $\delta_{r} \Sub(\Mbold)$.
In particular, since $\Tbold^{r} \in \delta_{r}\Sub(\Lbold)$,
$\delta_{r}\Vcal_{e}: \Lbold(e) \to \delta_{r} \Omegabold(e)$ 
which is defined by
\begin{equation}
\delta_{r}\Vcal_{e}:= \delta_{r}\chi^{\Tbold^{r} \Lbold}_{e}
\end{equation}
can be used as a valuation function alternative to $\Vcal^{r}_{e}$.
For any $e \in \Scal$ and $r \in \ES(R)$,
$\delta_{r}\Vcal_{e}$ satisfies the null-proposition condition.
In fact, $\delta_{r}\Vcal_{e}(\{ 0 \}) = \bot_{e_{r}}$,
which is the bottom of $\delta_{r} \Omegabold(e)$.

Although, for any $P \in  \Lcal$, $\delta_{r}\Vcal_{e}(P)=\Vcal^{r}_{e}(P)$
as a sieve on $e$,
they are different as truth-values because
the targets $\delta_{r} \Omega(e)$ and $\Omegabold(e)$
are different Heyting algebras.
We can, however, simply think of $\delta_{r}\Vcal_{e}$ 
as $\Vcal^{r}_{e}$ with a narrowed target $\delta_{r} \Omegabold(e)$.


\subsection{Alternative Construction of $\Vcal^{r}_{e}$}
\label{sebsec:Alternative Construction}

In the previous subsections,
the valuation functions $\Vcal^{r}_{e}$ are given
by the characteristic morphisms
$\chi^{\Tbold^{r} \Lbold}$
corresponding to the true subobjects $\Tbold^{r}$ of $\Lbold$.
We can, however, construct $\Vcal^{r}_{e}$
without using entire structure of $\Tbold^{r}$.
In fact, only the sets $\Tbold^{r}(e')$ such that 
$\Hom_{\CcalR}(e,e') \ne \emptyset $
are needed to define $\Vcal^{r}_{e}$.  
This suggests an alternative construction of $\Vcal^{r}_{e}$
by use of a restricted part of the base category.

For any state $e \in \Scal$,
we define a subcategory $\CcalRe$ of $\CcalR$;
its objects are given by
\begin{equation}
\Obj(\CcalRe) := \{ e' : \Hom_{\CcalR}(e,e') \ne \emptyset  \},
\end{equation}
and morphisms $\Mor(\CcalRe)$ are defined
for any $e'$ and $e'' \in \Obj(\CcalRe)$ by
\begin{equation}
\Hom_{\CcalRe}(e',e'') :=  \Hom_{\CcalR}(e',e'') .
\end{equation}
The subcategory $\CcalRe$ is, therefore, 
wide in $\CcalR$.

A restriction of a functor $\Kbold:\CcalR \to \Sets$ 
to the subcategory $\CcalRe$ (i.e., $\Kbold|_{\CcalRe}$) 
is denoted by $ \Kbold_{e \downarrow}$.
It is a functor from $\CcalRe$ to $\Sets$.
Note that, in particular, 
the functors ${\bold 1}_{e \downarrow}$ and $\Omegabold_{e \downarrow}$
are the final object and the subobject classifier 
in the topos of the functor category $\Sets^{\CcalRe}$,
respectively.

For any $e_{r} \in A(e)$ chosen as a true atom of $\Dcal(e,R)$,
we can define a global element of $\Abold_{e \downarrow }$,
$\sigma^{e_{r}}: {\bf 1}_{e \downarrow } \xrightarrow{\centerdot} \Abold_{ e \downarrow }$ by
$\sigma^{e_{r}}_{e}(*):= e_{r}$ with the naturality condition (\ref{eq:globalelement}).
This is nothing but a restriction of the global element $\sigma^{r}$ of $\Abold$
to the subcategory $\CcalRe$.

For each $e_{r} \in A(e)$,
we define a functor $ \Tbold^{e_{r}}  : \CcalRe \to \Sets$
by means of $\sigma^{e_r}$:
\begin{equation}
e'
\;
\longmapsto
\;
\Tbold^{e_{r}} (e') 
:= 
\{P \in \Lbold(e'): P \ge \sigma^{e_{r}}_{e'} (*)\},
\end{equation}
\begin{equation}
e' \xrightarrow{ \displaystyle \Fhat'} e''
\; \longmapsto \;
\begin{array}{ccc} \Tbold^{e_{r}} (e')  & \xrightarrow{\displaystyle  \Tbold^{e_{r}} (\Fhat')} & \Tbold^{e_{r}} (e'')  \\
[-1pt] \vertin &                 & \vertin \\
[-1pt] P  & \longmapsto   & \Fhat' P
\end{array}
\;.
\end{equation}
The functor $\Tbold^{e_{r}}$ is nothing but $\Tbold^{r}_{e \downarrow }$.

Since $\Tbold^{e_{r}}(e')$ is a filter for any $e' \in \Ocal(\CcalRe)$,
$\Tbold^{e_{r}}$ is a true subobject of $\Lbold_{e \downarrow }$
in the topos $\Sets^{\CcalRe}$.
Thus, the corresponding characteristic morphism 
 $\chi^{\Tbold^{e_r} \Lbold_{e \downarrow }} : 
\Lbold_{e \downarrow } \xrightarrow{\centerdot} \Omegabold_{e \downarrow }$
gives a truth-value for a proposition 
$P \in \Lcal =\Lbold_{e \downarrow }(e)$
at the stage $e$ by
\begin{eqnarray}
\chi^{\Tbold^{e_r} \Lbold_{e \downarrow }}_{e} (P) 
& = &
\bigcup\limits_{e' \in \Scal}\{ \Fhat \in \Hom_{\CcalRe} (e,e') : \Lbold_{e, \downarrow }(\Fhat)(P) \in  \Tbold^{e_r}(\Fhat)(e) \}
\nonumber \\
& = &
\bigcup\limits_{e' \in \Scal}\{ \Fhat \in \Hom_{\CcalRe} (e,e') : \Fhat(P) \in  \Tbold^{e_r}(e') \}
\nonumber \\
& = &
\bigcup\limits_{e' \in \Scal}\{ \Fhat \in \Hom_{\CcalRe} (e,e') : \Fhat(P) \geq \sigma^{e_{r}}_{e'}(*) = e'_{r} \}
\nonumber \\
& \in & 
\Omegabold_{e \downarrow }(e).
\label{eq:valuation3}
\end{eqnarray}
Equations (\ref{eq:valuation2}) and (\ref{eq:valuation3}) show that
$\chi^{\Tbold^{e_r} \Lbold_{e \downarrow }}_{e} (P) $ is equal to 
$\Vcal^{r}_{e}(P)$ as sets of morphisms.
Moreover, by the trivial correspondence,
$\Omegabold(e) \cong \Omegabold_{e \downarrow }(e)$ as Heyting algebras.
Therefore, $\chi^{\Tbold^{e_{r}} \Lbold_{e \downarrow }}_{e}$ gives
completely the same valuation as $\Vcal^{r}_{e}$.


\section{Valuations in Extended Functor Category}
\label{sec:Valuations in Extended}
\subsection{Construction of Extended Base Category $\Ccal$}
\label{subsec:Construction of Extended}
The purpose of Section \ref{sec:Valuations in Extended} is to construct valuation functions
in a topos which includes all determinate observables.
In order to do so,
in this subsection we construct a base category
which includes all observables.
First, we note that, if $\ES(R) = \ES(R')$ and $\Com(R) = \Com(R')$,
base categories $\CcalR$ and $\Ccal_{R'}$ give 
the same valuations for each state $e$.
We therefore deal with 
equivalent classes of observables consisting of 
those with the same eigenspaces and commutant.
To do so,
we define an order relation on the collection $\Ocal$ of the observables
by their functional relationship.
That is, for $R$, $R' \in \Ocal$,
\begin{equation}
R \le R' 
\;
\iff
\;
\exists f \in {\mathbb{R}}^{{\mathbb{R}}} \mbox{ s.t. } R = f(R'),
\end{equation}
and the equivalence relation is defined by
\begin{equation}
R \sim R' 
\;
\iff 
\;
R \le R', \mbox{  } R' \le R.
\end{equation}
The induced quotient space is denoted by $\Ocaltilde$.
It has natural order relation
induced by that of $\Ocal$;
that is, for any $\rho$, $\rho' \in \Ocaltilde$ such that
$\rho = [R]$ and $\rho' = [R']$,
\begin{equation}
\rho \le \rho' 
\;
\iff
\;
R \le R'
.
\end{equation}
The quotient space $\Ocaltilde$, thus, 
possesses a category structure
with the preorder relations as morphisms.
Furthermore, 
for any $\rho = [R]\in \Ocaltilde$,
the set of eigenspaces $\ES(\rho)$, 
the commutants $\Com(\rho)$, 
and the base category $\Ccal_{\rho}$ are given by
\begin{equation}
\ES(\rho) := \ES(R),
\mbox{  }
\Com(\rho) := \Com(R),
\mbox{  }
\Ccal_{\rho} := \CcalR,
\end{equation}
because the right-hand sides do not depend on which $R$ is chosen
as a representative of $\rho$.

We introduce category structures to the cartesian product $\Scal \times \Ocaltilde$.

First, we construct a category $\Bcal$ 
from the cartesian product $\Scal \times \Ocaltilde$
in such a way that each $\Ccal_{\rho}$ ($\rho \in \Ocaltilde$)
is naturally embedded.
That is,
the objects of $\Bcal$ are given by the elements of $\Scal \times \Ocaltilde$
and for any $(e,\rho)$, $(e', \rho') \in \Scal \times \Ocaltilde$
morphisms are given by
\begin{equation}
\Hom_{\Bcal} ((e,\rho),(e',\rho')) 
:=
\Hom_{\Ccal_{\rho}}(e,e') \times \Hom_{\Ocaltilde}(\rho,\rho').
\label{eq:morphism2}
\end{equation}
The composition of morphisms is defined as follows;
that is,  
for $(e \xrightarrow{\Fhat} e', \rho \leq \rho') \in \Hom_{\Bcal}((e,\rho),(e',\rho'))$
and
$(e' \xrightarrow{\Fhat} e'', \rho' \leq \rho'') \in \Hom_{\Bcal}((e',\rho'),(e'',\rho''))$,
\begin{eqnarray}
(e' \xrightarrow{\Fhat} e'', \rho' \leq \rho'') \circ (e \xrightarrow{\Fhat} e', \rho \leq \rho') 
& := & 
(e \xrightarrow{\Fhat'\Fhat} e'', \rho \leq \rho'')
\nonumber\\
& \in &
\Hom_{\Bcal}((e, \rho), (e'',\rho'')).
\label{eq:composition2}
\end{eqnarray}
Consistency of this composition rule is
ensured by definition (\ref{eq:morphism2})
and the fact that
\begin{equation}
\rho \le \rho' 
\;
\Longrightarrow
\;
\Com(\rho') \subseteq \Com(\rho).
\end{equation}
In fact,
$\Fhat' \in \Hom_{\Ccal_{\rho}}(e',e'')$ implies 
that $\Fhat' \in \Com(\rho') \subseteq \Com(\rho) $,
hence, for any $R \in \rho$,
$\Fhat'\Fhat \Rhat = \Fhat' \Rhat \Fhat = \Rhat \Fhat' \Fhat$;
that is, $\Fhat' \Fhat \in \Hom_{\Ccal_{\rho}} (e,e'')$.

For the sake of brevity,
we hereafter abbreviate the notation for morphisms
according to the following rule;
\begin{equation}
\Fhat \in \Hom_{\Bcal} ((e,\rho),(e',\rho')) 
\;
\iff
\;
\Fhat \in \Hom_{\Ccal_{\rho}} (e,e')
\mbox{ and }
\rho \leq \rho' .
\end{equation}
Also, we sometimes use another notation,
say, $\Fhat_{\rho \rho'}$
for $(e,\rho) \xrightarrow{\Fhat} (e',\rho')$,
in order to specify which objects of $\Ocaltilde$ occur as its domain and codomain. 

The category $\Bcal$ contains all of the categories 
$\Ccal_{\rho} \simeq \Scal \times \{ \rho \}$
 $(\rho \in \Ocaltilde)$ as subcategories.
It is, however,
not appropriate for a base category.
In the following, we see the reason.

The key ingredient in Section \ref{sec:Topos-Theoretic Valuations} 
is the functor $\Abold:\CcalR \to \Sets$.
Since any $R \in \rho$ has the same set of eigenspaces, $\ES(\rho)$,
the sets $A(e)$ of true atoms are also the same for each $e$.
Therefore, we can define a function ${\bar A} : \Obj(\Bcal) \rightarrow  \Obj(\Sets)$ by
\begin{equation}
{\bar A}(e,\rho) := \{ e_{r_1}, \cdots, e_{r_k} , \{0\} \} .
\end{equation} 
If we follow the line given in Section \ref{sec:Topos-Theoretic Valuations},
the function ${\bar A}$ should be extendable
to a functor $\Abold: \Bcal \rightarrow \Sets$
which maps $(e,\rho) \xrightarrow{\Fhat} (e',\rho)$
to a map $\Abold(\Fhat):{\bar A}(e,\rho) \to {\bar A}(e',\rho')$
defined by $\Abold(\Fhat)(e_{r}) := \Fhat(e_{r})$.
If $\rho \ne \rho'$, however,
in general $\Fhat(e_{r}) \not\in  A(e',\rho')$,
hence, ${\bar A}$ cannot define the functor $\Abold:\Bcal \to \Sets$.
In fact, since $\rho \leq \rho'$,
any eigenspace $r \in \ES(\rho)$
is uniquely decomposed by means of 
adequately chosen eigenspaces, $r'_{1}, \cdots , r'_{j}$,
of $\rho'$, as $r = r'_{1} \oplus  \cdots \oplus r'_{j} $.
If $\Fhat \in \Com(\rho)$, $\Fhat(r)$ is a subspace of $r$ 
but need not equal any one of $r'_{1}, \cdots , r'_{j} $.
Therefore $\Fhat(e_{r}) \in  A(e',\rho')$ cannot be concluded. 

In order to maintain the idea in Section \ref{sec:Topos-Theoretic Valuations} to the maximum extent possible,
we define a subcategory $\Ccal$
of $\Bcal$ in such a way that 
it contains all $\Ccal_{\rho}$ ($\rho \in \Ocaltilde$)
as subcategories
and that ${\bar A}(e,\rho)$ 
can be extended to a functor $\Abold:\Ccal \rightarrow \Sets$.
We can make good this 
by adopting $\Fhat \in \Hom_{\Bcal}((e,\rho), (e',\rho'))$ 
as a morphism of $\Ccal$ only if $\Fhat(e_{r}) \in  {\bar A}(e',\rho')$.
That is, 
the objects of $\Ccal$ are given by
\begin{equation}
\Obj(\Ccal):=\Obj(\Bcal)=\Scal \times \Ocaltilde,
\end{equation}
and the morphisms are defined by
\begin{equation}
\Hom_{\Ccal}((e,\rho),(e',\rho'))
:=
\{
\Fhat \in \Hom_{\Bcal}((e,\rho),(e',\rho')):
\;\Fhat ({\bar A}(e,\rho)) \subseteq {\bar A}(e',\rho')
\}.
\label{eq:morphism3}
\end{equation}
Here, note that, 
since $\Fhat ({\bar A}(e,\rho))={\bar A}(e',\rho)$,
the condition in the definition (\ref{eq:morphism3}) is equivalent to 
${\bar A}(e',\rho) \subseteq {\bar A}(e',\rho')$.
In fact, because of proposition B.1,
it is further reduced to ${\bar A}(e',\rho) = {\bar A}(e',\rho')$.

Since $\Fhat' \Fhat ({\bar A}(e,\rho)) \subseteq \Fhat'({\bar A}(e',\rho')) 
\subseteq {\bar A}(e'',\rho'')$
for any $\Fhat \in \Hom_{\Ccal}((e,\rho),(e',\rho'))$ and 
$\Fhat' \in \Hom_{\Ccal}((e',\rho'),(e'',\rho''))$,
definition (\ref{eq:morphism3}) of $\Mor(\Ccal)$ is 
consistent with the composition rule (\ref{eq:composition2}).

Note that, for any $\rho \in \Ocaltilde$, 
$\Ccal_{\rho}$ is a subcategory of $\Ccal$ as well as of $\Bcal$
because $\Ccal$ is a wide subcategory
and $\Hom_{\Ccal_{\rho}}(e,e') = \Hom_{\Ccal}((e,\rho),(e',\rho))$.
Further, any object in $\Ccal_{\rho}$ 
is connected to $\Ccal_{\rho'}$ with $\rho \leq \rho'$
via some morphisms.
In fact, for any $(e,\rho) \in \Obj(\Ccal)$ and $\rho' \ge \rho$,
there exists $e'$ s.t. $\Hom_{\Ccal}((e,\rho),(e',\rho')) \ne \emptyset $.
The simplest example is $e' = \pihat_{r'} e$ where 
$\pihat_{r'} \in \Hom_{\Ccal}((e,\rho),(e',\rho'))$ is
the projector on $r' \in \ES(\rho')$.

\subsection{Construction of Valuation Functions}
\label{Construction of Valuation}
The definition of the category $\Ccal$ allows
the map ${\bar A}$ to be a functor $\Abold:\Ccal \to \Sets$:
that is,
\begin{equation}
(e,\rho) 
\; \longmapsto \; 
\Abold(e,\rho) 
:=
{\bar A}(e,\rho)\;,
\end{equation}
\begin{equation}
(e,\rho) \xrightarrow{\displaystyle \Fhat} (e',\rho')
\; \longmapsto \;
\begin{array}{ccc} \Abold(e,\rho)  & \xrightarrow{\displaystyle \Abold( \Fhat)} & \Abold(e',\rho')  \\
[-1pt] \vertin &                 & \vertin \\
[-1pt] e_{r}  & \longmapsto   & \Fhat (e_{r}  )
\end{array}
\;
.
\end{equation}

Following Section \ref{sec:Topos-Theoretic Valuations},
we explicitly give definitions of
other functors to be needed.
The functor $\Lbold:\Ccal \rightarrow \Sets$,
which has $\Abold$ as a subobject,
is defined by
\begin{equation}
(e,\rho) 
\; \longmapsto \; 
\Lbold(e,\rho) 
:=
\Lcal
\;,
\end{equation}
\begin{equation}
(e,\rho) \xrightarrow{\displaystyle \Fhat} (e',\rho')
\; \longmapsto \;
\begin{array}{ccc} \Lbold(e,\rho)  & \xrightarrow{\displaystyle  \Lbold(\Fhat)} & \Lbold(e',\rho')  \\
[-1pt] \vertin &                 & \vertin \\
[-1pt] P  & \longmapsto   & \Fhat P
\end{array}
\;.
\end{equation}
The subobject classifier $\Omegabold$ of the topos $\Sets^{\Ccal}$ is given by
\begin{equation}
(e, \rho) 
\; \longmapsto \; 
\Omegabold(e,\rho) := \{ S: S \mbox{ is a sieve on } (e,\rho) \}
\;,
\end{equation}
\begin{equation}
(e,\rho) \xrightarrow{\displaystyle \Fhat} (e',\rho')
\; \longmapsto \;
\begin{array}{ccc} \Omegabold(e,\rho)  & \xrightarrow{\displaystyle  \Omegabold(\Fhat)} & \Omegabold(e',\rho')  \\
[-1pt] \vertin &                 & \vertin \\
[-1pt] S  & \longmapsto   & \left\{ \Fhat' \in \Mor(\Ccal) : \Fhat' \circ \Fhat\in S \right\}
\end{array}
\;.
\end{equation}
Also, the terminal object ${\bf 1}$ of $\Sets^{\Ccal}$ is 
defined by ${\bf 1}(e,\rho) := \{*\}$
and ${\bf 1}(\Fhat) := {\rm id}_{\{* \}}$.

Since $\Abold$ has no global elements,
we cannot go along the line in Section \ref{subsec:Prerequisites for True}.
To alter the way of construction, as in Section \ref{sebsec:Alternative Construction},
we define the full subcategory $\Ccal^{(e,\rho)\downarrow}$ 
of $\Ccal$ for any $(e,\rho) \in \Obj(\Ccal)$ by
\begin{equation}
\Obj(\Ccal^{(e,\rho)\downarrow})
:=
\{(e',\rho')\in \Obj(\Ocal):\Hom_{\Ccal}((e,\rho),(e',\rho')) \neq \emptyset \}
\end{equation}
and
\begin{equation}
\Hom_{\Ccal^{ (e,\rho)\downarrow }} ((e',\rho'),(e'',\rho'')) 
=
\Hom_{\Ccal} ((e',\rho'),(e'',\rho'')) .
\end{equation}

For any $(e,\rho) \in \Obj(\Ocal)$, 
the restriction $\Abold_{ (e,\rho)\downarrow }$ 
of $\Abold$ to $\Ccal^{(e,\rho)\downarrow }$
has a global element 
$\sigma^{e_{r}, \rho} : {\bf 1}_{ (e,\rho)\downarrow }
\xrightarrow{\centerdot} \Abold_{ (e,\rho)\downarrow }$
corresponding to each  $e_{r} \in \Abold(e,\rho)$.
This is uniquely determined by 
$\sigma^{e_{r}, \rho}_{(e,\rho)}(*):= e_{r} \in \Abold(e,\rho)$,
via the naturality condition applied 
to any $(e',\rho') \in \Ocal(\Ccal^{(e,\rho)\downarrow })$
and $\Fhat \in \Hom_{\Ccal}((e,\rho),(e',\rho'))$,
as 
\begin{equation}
\sigma^{e_{r}, \rho}_{(e',\rho')}(*) 
=
\Abold_{(e,\rho) \downarrow }(\Fhat)(\sigma^{e_{r}, \rho}_{(e,\rho)}(*))
=
\Fhat(e_{r})
=
e'_{r}
\in 
\Abold_{ (e,\rho)\downarrow }(e',\rho')
=
\Abold(e',\rho')
.
\end{equation}
It is easy to see that $\sigma^{e_{r}, \rho}$ is a natural transformation.

For any $(e,\rho) \in \Obj(\Ccal)$ and $ r \in \ES(\rho)$,
we define a functor $ \Tbold^{e_{r},\rho} : \Ccal^{(e,\rho)\downarrow } \to \Sets$
\begin{equation}
(e',\rho') 
\;
\longmapsto
\;
\Tbold^{e_{r},\rho} (e',\rho') 
:= 
\{P \in \Lbold_{(e,\rho)\downarrow }(e',\rho'): P \ge \sigma^{e_{r},\rho}_{(e',\rho')}(*)\}
\;,
\label{eq:truesubobject3a}
\end{equation}
\begin{equation}
(e',\rho') \xrightarrow{ \displaystyle \Fhat' } (e'',\rho'')
\; \longmapsto \; 
\begin{array}{ccc} \Tbold^{e_{r},\rho}  (e',\rho')  & \xrightarrow{\displaystyle  \Tbold^{e_{r},\rho}  (\Fhat')} & \Tbold^{e_{r},\rho}  (e'',\rho'')  \\
[-1pt] \vertin &                 & \vertin \\
[-1pt] P  & \longmapsto   & \Fhat' P
\end{array}
\; .
\label{eq:truesubobject3b}
\end{equation}
Since $\Tbold^{e_{r},\rho}$ is a filter,
its characteristic morphism
$\chi^{\Tbold^{e_{r},\rho}\, \Lbold_{(e,\rho) \downarrow }}:
{\bf 1}_{(e,\rho)\downarrow} \xrightarrow{\centerdot} \Omegabold_{(e,\rho)\downarrow}$ 
gives a truth-value valuation;
that is, for the truth atom $e_{r}$ at the stage $(e,\rho)$
and for any quantum proposition 
$P \in \Lbold_{(e,\rho) \downarrow }(e,\rho)=\Lbold(e,\rho) = \Lcal$,
we define the truth-value 
$\Vmath^{e_{r}, \rho}(P) \in \Omegabold_{(e,\rho)\downarrow }(e,\rho)= \Omegabold(e,\rho)$ by
\begin{eqnarray}
\Vmath^{e_{r}, \rho}(P)
& := &
\chi^{\Tbold^{e_{r},\rho} \Lbold_{(e,\rho) \downarrow }}_{(e,\rho)}(P) \nonumber \\
& = &
\{ 
\Fhat \in \Mor(\Ccal^{(e,\rho) \downarrow }):\dom(\Fhat)=(e,\rho), \;
\Lbold_{(e,\rho) \downarrow }(\Fhat)(P) \in  \Tbold^{e_{r},\rho}({\cod}(\Fhat))
\}
\nonumber \\
& = &
\bigcup\limits_{(e',\rho')\in \Obj(\Ccal^{(e,\rho)\downarrow })} 
\{ 
\Fhat \in \Hom_{\Ccal^{(e,\rho) \downarrow }} ((e,\rho),(e',\rho')):
\Fhat(P) \geq \sigma^{e_{r}, \rho}_{(e',\rho')} (*)
\} 
\nonumber \\
& = &
\bigcup\limits_{(e',\rho')\in \Obj(\Ccal^{(e,\rho)\downarrow })} 
\{ 
\Fhat \in \Hom_{\Ccal^{(e,\rho) \downarrow }} ((e,\rho),(e',\rho')):
\Fhat(P) \geq  e'_{r}
\} .
\end{eqnarray}
If there exists a filter subobject $\Tbold \in \Obj(\Sets^{\Ccal})$ of $\Lbold$
which satisfies $\Tbold_{\Ccal^{(e,\rho)\downarrow }} = \Tbold^{e_{r},\rho}$,
then it gives the same truth-value for each $P \in \Lcal$ 
at the stage $(e,\rho) \in \Obj(\Ccal)$.


\section{Relation between Valuation Structures Based on $\Sets^{\Ccal_{\rho}}$ and $\Sets^{\Ccal}$}
\label{sec:Relation between Valuation}
In the previous sections,
we constructed two types of valuation functions based on the toposes of presheaves,
one of which treated the case with fixed determinate observable
and the other formulated a framework in which all determinate observables 
are included.
They give, however, different results for the same situation;
for a state $e \in \Scal$ and a determinate observable $R \in \rho$, 
the assigned truth-values to a quantum proposition $P \in \Lcal$,
$\Vcal^{r}_{e}(P)$ and $\Vmath^{e_r, \rho}(P)$,
are different sieves on different Heyting algebras, 
$\Omegabold(e)$ and $\Omegabold(e,\rho)$, respectively.
In fact, 
$\Ccal$ includes morphisms between different $\rho$'s,
$\Vcal^{r}_{e}(P) \neq \Vmath^{e_r, \rho}(P)$ as sets of morphisms.
Also, $\Omegabold(e,\rho)$ includes sieves 
much more than $\Omegabold(e)$.
Therefore, neither structural relation between the Heyting algebras
nor logical relation between the truth-values is clear.
On the other hand,
definitions (\ref{eq:valuation2}) and (\ref{eq:valuation3}) of truth-value assignments
show that $\Vcal^{r}_{e}(P) = \Vmath^{e_r, \rho}(P)$
as the sets of linear operators of which domains and codomains are forgotten.
This suggests that 
there exists a Heyting algebra
on which $\Vcal^{r}_{e}(P)$ and $\Vmath^{e_r, \rho}(P)$ take the same value.
In this section,
we show that this is the case.

\subsection{Structural Relation between Subobject Classifiers
in $\Sets^{\Ccal_{\rho}}$ and $\Sets^{\Ccal}$}
\label{subsec:Structural Relation}
We investigate relation between
the subobject classifiers $\Omegabold_{\rho}$ of $\Sets^{\Ccal_{\rho}}$
and $\Omegabold$ of $\Sets^{\Ccal}$.
To do so, we define two important ingredients, $\sharp$ and $\flat$.
For any $(e,\rho) \in \Obj(\Ccal)$,
they give maps, $\sharp_{(e,\rho)}:\Omegabold_{\rho}(e) \rightarrow \Omegabold(e,\rho)$
and $\flat_{(e,\rho)}:\Omegabold(e,\rho) \rightarrow \Omegabold_{\rho}$,
respectively.
Their detailed definitions are described below.
Hereafter, the subscripts $(e,\rho)$ are omitted for the sake of brevity.

To define the map $\sharp:\Omegabold_{\rho}(e) \rightarrow \Omegabold(e,\rho)$,
we introduce $\eta:\Omegabold_{\rho}(e) \rightarrow \Sub(\Mor(\Ccal))$,
which lifts any sieve $S_{e} \in \Omegabold_{\rho}(e)$
to a set $\eta(S_{e})$ of morphisms in the category $\Ccal$ by 
\begin{eqnarray}
\eta(S_{e}) 
& := &
\left\{
\Ghat_{\rho \rho} \in \Mor(\Ccal): \Ghat \in S_{e} 
\right\}
\nonumber \\
& = & 
\bigcup\limits_{e'' \in \Scal}
\left\{
(e,\rho) \xrightarrow{\displaystyle \Ghat_{\rho\rho}} (e'', \rho) \in \Mor(\Ccal):
e \xrightarrow{\displaystyle \Ghat} e'' \in S_{e}
\right\}.
\end{eqnarray}
(As for the meaning of $\Ghat_{\rho \rho}$, see Section \ref{subsec:Construction of Extended}.)
Although $S_{e} \in \Omegabold_{\rho}(e)$, 
$\eta(S_{e}) \notin \Omegabold(e,\rho)$.
We define $\sharp(S_{e}) \in \Omegabold(e,\rho)$ as a
minimum sieve on $(e,\rho)$ including $\eta (S_{e})$;
that is,
for a family of sets, 
$
{\mathfrak U} := 
\{
S_{(e,\rho)} \in \Omegabold(e,\rho):
\eta (S_{e}) \subseteq S_{(e,\rho)}
\}
$,
$\sharp (S_{e})$ is defined by
$\sharp (S_{e}) := \bigcap {\mathfrak U}$
because any intersection of sieves on $(e,\rho)$
is also a sieve.
Also, it can be represented explicitly as 
\begin{equation}
\sharp(S_{e})
=
\left\{
\Fhat \in \Mor(\Ccal):
\exists \Ghat \in S_{e}
,\;
\exists \Hhat \in \Mor(\Ccal),
\mbox{ s.t. }
\Fhat = \Hhat \circ \Ghat_{\rho \rho}
\right\}.
\label{eq:sharp}
\end{equation}
In fact, it is easy to see that the right hand side of equation (\ref{eq:sharp})
is itself a sieve and is included by any sieve including $\eta(S_{e})$.

For any sieve $S_{(e,\rho)} \in \Omegabold(e,\rho)$,
we define a sieve $\flat(S_{(e,\rho)}) \in \Omegabold_{\rho}(e)$ by
\begin{equation}
\flat(S_{(e,\rho)})
:=
\left\{
\Fhat \in \Mor(\CcalR): \Fhat_{\rho \rho} \in S_{(e,\rho)}
\right\}.
\label{eq:flat}
\end{equation}

\begin{prp}
The map $\flat:\Omegabold(e,\rho)\rightarrow \Omegabold_{\rho}(e)$ is 
a lattice homomorphism preserving the top and the bottom.
\end{prp}
{\bf Proof.}
A lattice homomorphism is defined
as a join- and meet-preserving map.
To prove the join-preservation,
%
$
\flat(S_{1} \vee S_{2}) = \flat(S_{1}) \vee \flat(S_{2})
$,
%
suppose that $S_{1}$, $S_{2} \in \Omegabold(e,\rho)$.
Then we have the following equivalence relation:
\begin{eqnarray}
\Fhat \in \flat(S_{1} \vee S_{2}) 
& \iff &
\Fhat_{\rho \rho} \in S_{1} \vee S_{2}
\nonumber \\
& \iff &
\Fhat_{\rho \rho} \in S_{1} \mbox{ or } \Fhat_{\rho \rho} \in S_{2}
\nonumber \\
& \iff &
\Fhat \in \flat(S_{1}) \mbox{ or } \Fhat \in \flat(S_{2})
\nonumber \\
& \iff &
\Fhat \in \flat(S_{1}) \vee \flat(S_{2}) .
\label{eq:flatjoin}
\end{eqnarray}
We can prove the meet-preservation,
%
$
\flat(S_{1} \wedge  S_{2}) = \flat(S_{1}) \wedge  \flat(S_{2})
$,
%
by replacing 
the symbol $\vee$ and the word `or' by $\wedge$ and `and', respectively,
in (\ref{eq:flatjoin}).

The top- and bottom-preservation,
$\flat(\top_{(e,\rho)}) = \top_{e}$,
and
$\flat(\bot_{(e,\rho)}) = \bot_{e}$,
%
are clear from the definition (\ref{eq:flat}).
\hfill\qed

\begin{prp}
The map $\sharp:\Omegabold_{\rho}(e)\rightarrow \Omegabold(e,\rho)$ is a lattice homomorphism
preserving the top and the bottom.
\end{prp}
{\bf Proof.}
Preservation of $\top$ and $\bot $ is clear.
To see the $\vee $-preservation,
let $S_{1}$, $S_{2} \in \Omegabold_{\rho}(e)$,
then we have
\begin{eqnarray}
\Fhat \in \sharp(S_{1} \vee S_{2})
& \iff &
\exists \Ghat \in S_{1} \vee S_{2},
\;
\exists \Hhat \in \Mor(\Ccal),
\mbox{ s.t. }
\Fhat = \Hhat \circ \Ghat_{\rho \rho}
\nonumber \\
& \iff &
\exists \Ghat \in \Mor(\CcalR),
\;
\exists \Hhat \in \Mor(\Ccal),
\nonumber \\
& \mbox{ } &
\qquad
\mbox{ s.t. }
\Ghat \in S_{1} \mbox{ or } \Ghat \in S_{2},\;
\Fhat = \Hhat \circ \Ghat_{\rho \rho}
\nonumber \\
& \iff &
\Fhat \in \sharp(S_{1}) \vee \sharp(S_{2}).
\end{eqnarray}
Similarly, we can verify the $ \wedge $-preservation.
\hfill\qed

In general, lattice homomorphisms automatically 
satisfy the order-preservation (e.g.,~\cite{DP90}).
Therefore, $\flat:\Omegabold(e,\rho) \to \Omegabold_{\rho}(e)$ 
and $\sharp:\Omegabold_{\rho}(e) \to \Omegabold(e,\rho)$ 
are order homomorphisms.
(Of course, this can be directly verified 
from the definitions (\ref{eq:sharp}) and (\ref{eq:flat}).)
They are not, however, Heyting-algebra homomorphisms,
because they do not preserve the pseudocomplements ($\Rightarrow $).
In fact, what we can safely say about their effects 
on the pseudocomplements is the following order relation:
\begin{equation}
f(S_{1} \Rightarrow S_{2}) \leq (f(S_{1}) \Rightarrow f(S_{2})),
\label{eq:fpseudocomplement}
\end{equation}
where $f = \sharp$ or $\flat$.
Inequality (\ref{eq:fpseudocomplement}) results from the fact that 
the maps are lattice homomorphisms.
In fact, if $S$, $S_{1}$, $S_{2} \in \Omegabold(e,\rho)$,
then
\begin{eqnarray}
S \wedge S_{1} \leq S_{2}
 &\Longrightarrow & 
f(S \wedge S_{1}) \leq f(S_{2}) \nonumber\\
 &\iff  & 
f(S) \wedge f(S_{1}) \leq f(S_{2}) \nonumber\\
&\Longrightarrow  & 
f(S) \leq (f(S_{1}) \Rightarrow  f(S_{2})).
\end{eqnarray}
In particular,
letting $S$ be $S_{1} \Rightarrow  S_{2}$,
we have an inequality which is always true, 
$(S_{1} \Rightarrow  S_{2}) \wedge S_{1} \leq S_{2}$,
as the leftmost inequality,
hence, inequality (\ref{eq:fpseudocomplement}) holds.

\begin{prp}
The composition of lattice homomorphisms, 
$\flat \circ \sharp:\Omegabold_{\rho}(e)\to \Omegabold_{\rho}(e)$ satisfy the equality,
\begin{equation}
\flat \circ \sharp = {\rm 1}_{\Omegabold_{\rho}(e)}.
\label{eq:flatsharp}
\end{equation}
Thus, the maps $\sharp$ is injective and $\flat$ surjective.
On the other hand, 
$\natural := \sharp \circ \flat:\Omegabold(e,\rho) \to \Omegabold(e,\rho)$ 
is a lattice homomorphism preserving the top and the bottom,
and satisfies the inequality,
\begin{equation}
\natural := \sharp \circ \flat \leq  {\rm 1}_{\Omegabold(e,\rho)}.
\label{eq:natural}
\end{equation}
\end{prp}
{\bf Proof.}
For any $S_{e} \in \Omegabold_{\rho}(e)$
\begin{eqnarray}
\Fhat \in \flat \circ \sharp (S_{e})
& \iff &
\Fhat_{\rho \rho} \in \sharp(S_{e})
\nonumber \\
& \iff &
\exists \Ghat \in S_{e}, \; \exists \Hhat \in \Mor(\Ccal),
\mbox{ s.t. }
\Fhat_{\rho \rho} = \Hhat \circ \Ghat_{\rho \rho}\,.
\end{eqnarray}
This implies that, in $\CcalR$, 
$\Fhat = \Hhat \circ \Ghat$.
Thus, since $\Ghat \in S_{e}$, $\Fhat \in S_{e}$.
Conversely,
$ \Fhat \in S_{e} $
implies that $\Fhat_{\rho \rho} \in \sharp (S_{e})$, hence, $\Fhat \in b(\sharp(S_{e}))$.
Thus, equality (\ref{eq:flatsharp}) is verified.

Since $\flat$ and $\sharp$ are lattice homomorphisms preserving 
the top and the bottom,
so is $\natural$.
To show inequality (\ref{eq:natural}),
suppose that $\Fhat \in \sharp \circ \flat (S_{(e,\rho)})$.
Then there exist $\Ghat \in \flat(S_{(e,\rho)})$ and $\Hhat \in \Mor(\Ccal)$ 
such that $\Fhat = \Hhat \circ \Ghat_{\rho \rho}$.
But then, because of the definition of $\flat$, 
$\Ghat_{\rho \rho} \in S_{(e,\rho)}$.
Thus, $\Fhat \in S_{(e,\rho)}$ because $S_{(e,\rho)}$ is a sieve.
\hfill\qed

With regard to the lattice-homomorphism 
$\natural:\Omegabold(e,\rho) \to \Omegabold(e,\rho)$,
we introduce the following definition:
\begin{dfn}
A sieve $S \in \Omegabold(e,\rho)$ is said to be natural
if it is a fixpoint of $\natural$,
i.e., $\natural (S) = S$.
\end{dfn}

\begin{prp}
A sieve $S \in \Omegabold(e,\rho)$ is natural
if and only if $S \in \natural(\Omegabold(e,\rho))$.
That is, $\natural(\Omegabold(e,\rho))$ is a set of 
fixpoints of $\natural$.
\end{prp}
{\bf Proof.}
If $S$ is natural, $S = \natural(S)$, hence $S \in \natural(\Omegabold(e,\rho))$.
Conversely, if $S \in \natural(\Omegabold(e,\rho))$,
there exists $S' \in \Omegabold(e,\rho)$ s.t. $S = \natural(S')$.
Then, $\natural (S) = \natural \circ \natural(S') 
= \sharp \circ \flat \circ \sharp \circ \flat (S') = \sharp \circ \flat (S') = \natural (S') = S$.
Thus, $S$ is natural.
\hfill\qed

Propositions 53 and 55 imply 
that the restriction $\flat'$ of the map $\flat$ to $\natural(\Omegabold(e,\rho))$,
$\flat':\natural(\Omegabold(e,\rho)) \to \Omegabold_{\rho}(e)$,
and $\sharp:\Omegabold_{\rho}(e) \to \natural(\Omegabold(e,\rho))$
are mutually inverse;
they are lattice-isomorphisms between
$\natural(\Omegabold(e,\rho))$ and $\Omegabold_{\rho}(e)$.
Moreover, $\natural(\Omegabold(e,\rho))$ is a Heyting-algebra
isomorphic to $\Omegabold_{\rho}(e)$,
because for any $S_{1}$, $S_{2} \in \natural(\Omegabold(e,\rho))$,
their pseudocomplement 
$S_{1} \Rightarrow_{\natural} S_{2} \in \natural(\Omegabold(e,\rho))$ 
can be defined by
\begin{equation}
S_{1} \Rightarrow_{\natural} S_{2}
:=
\sharp(\flat(S_{1}) \Rightarrow \flat(S_{2})).
\end{equation}
The map $\flat':\natural(\Omegabold(e,\rho)) \to \Omegabold_{\rho}(e)$
is, therefore, a Heyting-algebra isomorphism.

We, thus, obtain the following theorem:
\begin{thm}
For any $(e,\rho) \in \Obj(\Ccal)$,
the Heyting algebra $\Omegabold(e,\rho)$ includes a sublattice
$\natural (\Omegabold(e,\rho))= \sharp(\Omegabold_{\rho}(e))$
equipped with the top $\top_{(e,\rho)}$ and the bottom $\bot_{(e,\rho)}$.
It is also a Heyting algebra isomorphic to $\Omegabold_{\rho}(e)$.
\end{thm} 

This theorem does not mean that
$\natural (\Omegabold(e,\rho))$ is a Heyting subalgebra
of $\Omegabold(e,\rho)$.
In general, 
we cannot assert that $S_{1} \Rightarrow_{\natural} S_{2}$ should equal
$S_{1} \Rightarrow S_{2}$ defined on $\Omegabold(e,\rho)$.
In fact, we can only say that 
\begin{eqnarray}
S_{1} \Rightarrow_{\natural} S_{2}
 & = & 
\sharp(\flat(S_{1}) \Rightarrow \flat(S_{2})) \nonumber\\
 & \leq  & 
\sharp(\flat(S_{1})) \Rightarrow \sharp(\flat(S_{2})) \nonumber\\
 & = & 
S_{1} \Rightarrow S_{2}.
\end{eqnarray}
Not only that, even if $S_{1}$ and $S_{2}$ belong to $\natural (\Omegabold(e,\rho))$,
$S_{1} \Rightarrow S_{2}$ need not belong to  $\natural (\Omegabold(e,\rho))$
for $(e,\rho) \in \Obj(\Ccal)$.

As is seen below, further,
the sets $\natural (\Omegabold(e,\rho))$ make up a functor.

\begin{prp}
If $S \in \Omegabold(e,\rho)$ is natural,
then so is $\Omegabold(\Fhat)(S) \in \Omegabold(e',\rho')$
for any $\Fhat \in \Hom_{\Ccal}((e,\rho),(e',\rho'))$. 
\end{prp}
{\bf Proof.}
Since we have inequality ($\ref{eq:natural}$), 
in order to prove that $\Omegabold(\Fhat)(S)$ is natural,
it suffices to show that
$\Omegabold(\Fhat)(S) \leq  \natural (\Omegabold(\Fhat)(S))$
for any natural $S \in \Omegabold(e,\rho)$.

To do so, 
suppose that $(e',\rho') \xrightarrow{\Fhat'} (e'',\rho'') \in \Omegabold(\Fhat) (S)$.
Then we have
$
(e,\rho) \xrightarrow{\Fhat}(e',\rho') 
\xrightarrow{\Fhat'} (e'',\rho'') \in S
$.
Since $S$ is natural, i.e., $S = \sharp(\flat(S))$,
there exist arrows, $e \xrightarrow{\Ghat} e''' \in \flat(S)$ and 
$\Hhat_{\rho \rho''} \in \Hom_{\Ccal}((e''',\rho),(e'',\rho''))$,
such that the following diagram commutes:
\begin{equation}
\begin{CD}
(e,\rho) @ > \Fhat_{\rho \rho'} >> (e', \rho') \\
   @ V \Ghat_{\rho \rho} VV @ VV \Fhat'_{\rho' \rho''} V \\
(e''',\rho) @ >> \Hhat_{\rho \rho''} > (e'',\rho'')
\end{CD}
\;\;\; .
\label{eq:F'F=HG1}
\end{equation}

For the operator $\Hhat$ corresponding to $\Hhat_{\rho \rho''}$,
we have $\Hhat \in \Hom_{\Ccal}((e''',\rho),(e'',\rho))$,
hence, $\Hhat \in \Hom_{\Ccal}((e''',\rho),(e'',\rho'))$
because of (\ref{eq:morphism3}) and proposition B.2.
Thus, we obtain the following commutative diagram:
\begin{equation}
\begin{CD}
 (e''',\rho) @ = (e''', \rho) @ =  (e''', \rho)     \\
@ V \Hhat_{\rho \rho} VV  @ VV \Hhat_{\rho \rho'} V  @ VV \Hhat_{\rho \rho''} V \\
(e'',\rho) @ >> \Ihat_{\rho \rho'} > (e'',\rho') @ >> \Ihat_{\rho' \rho''} > (e'', \rho'')
\end{CD}
\;\;\;.
\end{equation}

On the other hand, we have $\Fhat' \in \Hom_{\Ccal}((e',\rho'),(e'',\rho'))$ 
for $\Fhat' = \Fhat'_{\rho' \rho''}$.
Also, it holds that $\Fhat' \Fhat = \Hhat \Ghat$ as linear transformations of $\Hcal$
because of commutative diagram (\ref{eq:F'F=HG1}).
Thus, the following diagram commutes:
\begin{equation}
\begin{CD}
(e,\rho) @ > \Fhat_{\rho \rho'} >> (e', \rho') \\
   @ V \Ghat_{\rho \rho} VV @ VV \Fhat'_{\rho' \rho'} V \\
(e''',\rho) @ >> \Hhat_{\rho \rho'} > (e'',\rho')
\end{CD}
\quad .
\label{eq:F'F=HG2}
\end{equation}

Since $\Ghat \in \flat(S)$, we have $\Ghat_{\rho \rho} \in S$.
Therefore the commutative diagram (\ref{eq:F'F=HG2}) means that 
$\Fhat'_{\rho' \rho'} \circ \Fhat_{\rho \rho'} \in S$,
which implies $\Fhat'_{\rho' \rho'} \in \Omegabold(\Fhat)(S)$,
hence, $e' \xrightarrow{\Fhat'} e'' \in \flat(\Omegabold(\Fhat)(S))$.

Finally, note that the diagram
\begin{equation}
\begin{CD}
(e',\rho') @ > \Fhat'_{\rho' \rho''} >> (e'', \rho'') \\
   @ V \Fhat'_{\rho' \rho'} VV @ | \\
(e'',\rho') @ >> \Ihat_{\rho' \rho''} > (e'',\rho'')
\end{CD}
\;\;
\end{equation}
commutes.
This diagram implies that 
$(e',\rho') \xrightarrow{\Fhat'} (e'',\rho'') 
\in  \natural (\Omegabold(\Fhat)(S))$
because $\Fhat' \in \flat(\Omegabold(\Fhat)(S))$ as is shown above.
\hfill\qed

Theorem 56 and Proposition 57 entail the following theorem:
\begin{thm}
The sets $\natural(\Omegabold(e,\rho))$ ($(e,\rho) \in \Obj(\Ccal)$),
each of which is a set of natural sieves on $(e,\rho)$,
can be extended to a functor from $\Ccal$ to $\Sets$,
which is hereafter denoted by $\natural \Omegabold$:
\begin{equation}
(e,\rho) 
\; \longmapsto \; 
\natural\Omegabold(e,\rho) 
:=
\natural(\Omegabold(e,\rho)) 
=\{S \in \Omegabold(e,\rho): \natural (S) = S\},
\end{equation}
\begin{equation}
(e,\rho) \xrightarrow{\displaystyle \Fhat} (e',\rho')
\; \longmapsto \;
\begin{array}{ccc} \natural \Omegabold (e,\rho)  & \xrightarrow{\displaystyle \natural \Omegabold ( \Fhat)} & \natural \Omegabold (e',\rho')  \\
[-1pt] \vertin &                 & \vertin \\
[-1pt] S  & \longmapsto   & \Omegabold(\Fhat) (S)
\end{array}
\; .
\end{equation}
\end{thm}
Note that the functor $\natural \Omegabold$ is a subobject 
of the subobject classifier $\Omegabold$.
Furthermore, as is seen in the next subsection,
the functor $\natural \Omegabold$ behaves as a subobject classifier
for a particular collection of subobjects of each functors;
that is, $\natural \Omegabold$ is a subobject semi-classifier
of the subobjects.

\subsection{Logical Relation between Valuations 
in $\Sets^{\Ccal_{\rho}}$ and $\Sets^{\Ccal}$}
\label{subsec:Logical Relation}
The definition of the functor $\Tbold^{e_{r},\rho}$, (\ref{eq:truesubobject3a}) and (\ref{eq:truesubobject3b}), 
 entails that,
for any proposition $P \in \Lbold_{(e,\rho)\downarrow }(e',\rho')$ and 
any $(e',\rho') \xrightarrow{\Fhat'} (e'',\rho'') \in \Mor(\Ccal^{(e,\rho)\downarrow })$,
if $\Lbold_{(e,\rho)\downarrow }(\Fhat'_{\rho' \rho''})(P) \in \Tbold^{e_{r},\rho}(e'',\rho'')$,
then $\Lbold_{(e,\rho)\downarrow }(\Fhat'_{\rho' \rho'})(P) \in \Tbold^{e_{r},\rho}(e'',\rho')$.
In fact, definition (\ref{eq:truesubobject3a}) gives the following 
equivalence relation:
\begin{eqnarray}
\Lbold_{(e,\rho)\downarrow }(\Fhat'_{\rho' \rho''})(P) \in \Tbold^{e_{r},\rho}(e'',\rho'')
& \iff &
\Fhat'(P) \geq \sigma^{e_{r}, \rho}_{(e'',\rho'')}(*)
\nonumber \\
& \iff &
\Fhat'(P) \geq \Fhat'\Fhat(e_{r})
\nonumber \\
& \iff &
\Fhat'(P) \geq \sigma^{e_{r}, \rho}_{(e'',\rho')}(*)
\nonumber \\
& \iff &
\Lbold_{(e,\rho)\downarrow }(\Fhat'_{\rho' \rho'})(P) \in \Tbold^{e_{r},\rho}(e'',\rho'),
\nonumber \\
& \mbox{ } &
\mbox{ }
\end{eqnarray}
where the operator $\Fhat$ occurring 
in the second line is an arbitrary morphism contained by 
$\Hom_{\Ccal^{(e,\rho)\downarrow }}((e,\rho),(e',\rho'))$.
We generalize the above-mentioned property as follows:
\begin{dfn}
Let $\Mbold$ be an object of $\Obj(\Sets^{\Ccal})$.
A subobject $\Nbold$ of $\Mbold$ 
is said to be projective for $x \in \Mbold(e,\rho)$
if, for any $\Fhat \in \Hom_{\Ccal}((e,\rho),(e',\rho'))$,
\begin{equation}
\Mbold(\Fhat_{\rho \rho'})(x) \in \Nbold(e',\rho')
\;
\Longrightarrow 
\;
\Mbold(\Fhat_{\rho \rho})(x) \in \Nbold(e',\rho).
\label{eq:O-connected}
\end{equation} 
If the implication relation (\ref{eq:O-connected}) holds
for all $x \in \Mbold(e,\rho)$,
$\Nbold$ is said to be projective for $\Mbold(e,\rho)$.
Also, if it does for all $(e,\rho) \in \Obj(\Ccal)$,
$\Nbold$ is said to be projective for $\Mbold$. 
\end{dfn}
Note that we do not propose the converse of (\ref{eq:O-connected})
because the left-hand side
immediately results from the right-hand side.

\begin{prp}
The subobject $\Nbold$ of $\Mbold$ is projective for $x \in \Mbold(e,\rho)$
if and only if 
$\chi^{\Nbold \Mbold}_{(e,\rho)}(x) \in \Omega(e,\rho)$ is natural.
\end{prp}
{\bf Proof.}
($\Longrightarrow$)
Suppose that $\Nbold$ is projective for $x \in \Mbold(e,\rho)$.
Then, we have the implication relation,
\begin{eqnarray}
(e,\rho) \xrightarrow{\Fhat} (e',\rho') \in \chi^{\Nbold \Mbold}_{(e,\rho)}(x)
& \iff &
\Mbold(\Fhat_{\rho \rho'})(x) \in \Nbold(e',\rho') \nonumber \\
 & \Longrightarrow  & \;
\Mbold(\Fhat_{\rho \rho})(x) \in \Nbold(e',\rho) \nonumber \\
& \iff &
(e,\rho) \xrightarrow{\Fhat} (e',\rho) \in \chi^{\Nbold \Mbold}_{(e,\rho)}(x) \nonumber\\
& \iff &
e \xrightarrow{\Fhat} e' \in \flat (\chi^{\Nbold \Mbold}_{(e,\rho)}(x)) \nonumber\\
& \Longrightarrow &
(e,\rho) \xrightarrow{\Fhat} (e',\rho') 
\in \natural(\chi^{\Nbold \Mbold}_{(e,\rho)}(x)) ,
\end{eqnarray}
which means that 
$\chi^{\Nbold \Mbold}_{(e,\rho)}(x) \leq  \natural (\chi^{\Nbold \Mbold}_{(e,\rho)}(x))$,
hence, $\chi^{\Nbold \Mbold}_{e,\rho}(x)$ is natural.
[In the relation (5.21), the last line comes from 
$\Fhat_{\rho \rho'} = \Ihat_{\rho \rho'} \circ  \Fhat_{\rho \rho}$.]

($\Longleftarrow$)
Suppose that $\chi^{\Nbold \Mbold}_{(e,\rho)}(x)$ is natural,
i.e., 
$\chi^{\Nbold \Mbold}_{(e,\rho)}(x) =
\natural(\chi^{\Nbold \Mbold}_{(e,\rho)}(x))$.
Then, 
\begin{eqnarray}
\Mbold(\Fhat_{\rho \rho'})(x) \in \Nbold(e',\rho')
 & \iff & 
\Fhat_{\rho \rho'} \in \chi^{\Nbold \Mbold}_{(e,\rho)}(x) 
\nonumber \\
 & \Longrightarrow & 
\Fhat_{\rho \rho} \in \chi^{\Nbold \Mbold}_{(e,\rho)}(x)
\nonumber \\
 & \Longleftrightarrow & 
\Mbold(\Fhat_{\rho \rho})(x) \in \Nbold(e',\rho),
\end{eqnarray}
where,
the second line comes from the assumption
$ \chi^{\Nbold \Mbold}_{(e,\rho)}(x) 
= \natural(\chi^{\Nbold \Mbold}_{(e,\rho)}(x))$
and Proposition C1 in Appendix C. 
\hfill\qed

\begin{cor}
If $\Nbold \hookrightarrow \Mbold$ is projective for $\Mbold(e,\rho)$,
then the equation for $\Sets$-morphisms,
\begin{equation}
\chi^{\Nbold \Mbold}_{(e,\rho)} 
= 
\iota^{\natural \Omegabold \, \Omegabold}_{(e,\rho)} 
\circ \natural \circ \chi^{\Nbold \Mbold}_{(e,\rho)},
\label{eq:iota-natural-chi1}
\end{equation}
follows.
\end{cor}

\begin{thm}
If $\Nbold \hookrightarrow \Mbold$ is projective for $\Mbold$,
the functions $\natural \circ \chi^{\Nbold \Mbold}_{(e,\rho)}:
\Mbold(e,\rho) \rightarrow \natural \Omegabold (e,\rho)$ 
($(e,\rho) \in \Obj(\Ccal)$) defines a natural transformation,
which is hereafter denoted by 
$\natural \chi^{\Nbold \Mbold}:\Mbold \xrightarrow{\centerdot} \natural \Omegabold$.
Further, the following equation for the natural transformations holds:
\begin{equation}
\chi^{\Nbold \Mbold} 
=
\iota^{\natural \Omegabold \, \Omegabold} \circ \natural \chi^{\Nbold \Mbold}.
\label{eq:iota-natural-chi2}
\end{equation}
\end{thm}
{\bf Proof.}
From Theorems 58, for any natural $S \in \Omegabold(e,\rho)$
and $(e,\rho) \xrightarrow{\Fhat} (e',\rho') \in \Mor(\Ccal)$, 
it follows that 
\begin{equation}
\natural (\Omegabold(\Fhat)(S)) = \Omegabold(\Fhat)(S) =\Omegabold(\Fhat)(\natural(S)).
\end{equation}
Therefore, the diagram
\begin{equation}
\begin{CD}
 \Mbold(e,\rho) @ > \chi^{\Nbold \Mbold}_{(e,\rho)} >> \Omegabold(e, \rho) @ > \natural_{(e,\rho)} >>  \natural \Omegabold(e, \rho)    \\
@.  @ VV \Omegabold(\Fhat) V  @ VV \natural \Omegabold(\Fhat)  V \\
 @.  \Omegabold(e', \rho') @ >> \natural_{(e',\rho')} >  \natural \Omegabold(e', \rho')
\end{CD}
\label{eq:MOmegaOmega1}
\end{equation}
commutes for any $(e,\rho) \xrightarrow{\Fhat} (e', \rho')$
because, for any $x \in \Mbold(e,\rho)$,
$\chi^{\Nbold \Mbold}_{(e,\rho)}(x)$ is natural
from Proposition 510.

The commutativity of diagram (\ref{eq:MOmegaOmega1}) and
the naturality of $\chi^{\Nbold \Mbold}: \Mbold \xrightarrow{\centerdot} \Omegabold$
further ensure that the outer square of the diagram
\begin{equation}
\begin{CD}
 \Mbold(e,\rho) @ > \chi^{\Nbold \Mbold}_{(e,\rho)} >> \Omegabold(e, \rho) @ > \natural_{(e,\rho)} >>  \natural \Omegabold(e, \rho)    \\
@ V \Mbold(\Fhat) VV  @ VV \Omegabold(\Fhat) V  @ VV \natural \Omegabold(\Fhat)  V \\
\Mbold(e',\rho') @ >> \chi^{\Nbold \Mbold}_{(e',\rho')} > \Omegabold(e', \rho') @ >> \natural_{(e',\rho')} >  \natural \Omegabold(e', \rho')
\end{CD}
\end{equation}
commutes.
This shows naturality of $\natural \chi^{\Nbold \Mbold}$.
Equation (\ref{eq:iota-natural-chi2}) is a straightforward result 
from equation (\ref{eq:iota-natural-chi1}) which holds object-wise. 
\hfill\qed

To present main theorems,
we introduce a morphism {\it true} into $\natural \Omegabold$,
i.e., a natural transformation 
$\natural \tau : {\bf 1} \xrightarrow{\centerdot} \natural \Omegabold$,
which is defined by $\natural \tau_{(e,\rho)}(*) = \top_{(e,\rho)} \in \natural \Omegabold(e,\rho)$
for each $(e,\rho) \in \Obj(\Ccal)$.
Note that $\natural \tau$ can be also defined as a pullback of the {\it true},
$\tau : {\bf 1} \xrightarrow{\centerdot} \Omegabold$, 
along the inclusion morphism 
$\iota^{\natural \Omegabold \, \Omegabold} : 
\natural \Omegabold \xrightarrow{\centerdot} \Omegabold$;
that is, the diagram
\begin{equation}
\begin{CD}
{\bf 1} @ = {\bf 1} \\
   @ V {\natural \tau} VV @ VV  \tau V \\
 \natural \Omegabold @ >>  \iota^{\natural\Omegabold \, \Omegabold} >  \Omegabold
\end{CD}
\end{equation}
is a pullback.
Thus, we can apply Propositions A1 and A2:
\begin{thm}
If $\Nbold$ is a projective subobject for $\Mbold \in \Obj(\Sets^{\Ccal })$,
the diagram
\begin{equation}
\begin{CD}
\Nbold @ >  ! >> {\bf 1} \\
   @ V {\iota^{\Nbold \Mbold}} VV @ VV  \natural \tau V \\
 \Mbold @ >> \natural \chi^{\Nbold \Mbold} > \natural \Omegabold
\end{CD}
\end{equation}
is a pullback.
\end{thm}

More precisely,
we can show that one-to-one correspondence between 
classes of isomorphic projective subobjects of $\Mbold$ 
and natural transformations from $\Mbold$ to $\natural \Omegabold$.  
\begin{thm}
(i) Suppose that $\Nbold \hookrightarrow \Mbold$ is projective for $\Mbold$.
If there exists a $\Sets^{\Ccal}$-morphism 
$\zeta: \Mbold \xrightarrow{\centerdot} \natural \Omegabold$
which makes the diagram 
\begin{equation}
\begin{CD}
\Nbold @ >  ! >> {\bf 1} \\
   @ V {\iota^{\Nbold \Mbold}} VV @ VV \natural \tau V \\
 \Mbold @ >>  \zeta> \natural \Omegabold
\end{CD}
\label{eq:NMzetanaturalOmega}
\end{equation}
a pullback,
then $\zeta = \natural \chi^{\Nbold \Mbold}$.

(ii) Conversely, for each $\Sets^{\Ccal}$-morphism
$\zeta: \Mbold \xrightarrow{\centerdot} \natural \Omegabold$,
there exist projective subobjects $\Nbold$ of $\Mbold$,
being determined up to isomorphism,
such that $\zeta = \natural \chi^{\Nbold \Mbold}$,
and hence, the corresponding diagram (\ref{eq:NMzetanaturalOmega}) becomes a pullback.
\end{thm}

Next, we describe how the projective subobjects in $\Sets^{\Ccal}$ relate to
$\Sets^{\Ccal_{\rho}}$.

Note that, for each object $\Mbold$ of $\Sets^{\Ccal}$
and each $\rho \in \Obj(\Ccal)$, 
we can define an object $\Mbold|_{\rho}$ 
of $\Sets^{\Ccal_{\rho}}$
by $\Mbold|_{\rho} (e) := \Mbold(e,\rho)$
and $\Mbold(e \xrightarrow{\Fhat} e') 
:= \Mbold((e,\rho) \xrightarrow{\Fhat} (e',\rho))$.
We have the following proposition:
\begin{prp}
Suppose that $\Mbold$ is an object of $\Sets^{\Ccal}$ 
and $\Nbold$ is a subobject of $\Mbold$.
Then, for each $(e,\rho)$, the diagram
\begin{equation}
\begin{CD}
\Mbold(e,\rho) @ > \chi^{\Nbold \Mbold}_{(e,\rho)} >> \Omegabold(e,\rho)\\
   @ V i_{(e,\rho)} VV @ VV \flat V \\
\Mbold|_{\rho}(e)  @ >> \chi^{\Nbold|_{\rho}\, \Mbold|_{\rho}}_{e} > \Omegabold_{\rho}(e)
\end{CD}
\; \; 
\label{eq:logicalrelation1}
\end{equation}
commutes,
where $i_{(e,\rho)}: \Mbold(e,\rho) \xrightarrow{\sim } \Mbold|_{\rho}(e)$
is the trivial map $x \mapsto x$.
Further, if $\Nbold$ is projective for $\Mbold$, 
the diagram (\ref{eq:logicalrelation1}) is partitioned into 
two commutative squares:
\begin{equation}
\begin{CD}
\Mbold(e,\rho) @ > \natural \chi^{\Nbold \Mbold}_{(e,\rho)} >> 
\natural\Omegabold(e,\rho) @ > \iota^{\natural\Omegabold \, \Omegabold}_{(e,\rho)} >> 
\Omegabold(e,\rho) \\
   @ V i_{(e,\rho)} VV @ VV \flat' V  @ VV \flat V \\
\Mbold|_{\rho}(e)  @ >> \chi^{\Nbold|_{\rho}\, \Mbold|_{\rho}}_{e} > 
\Omegabold_{\rho}(e) @ = \Omegabold_{\rho}(e)
\end{CD}
\; \; .
\label{eq:logicalrelation2}
\end{equation}
\end{prp}
{\bf Proof.}
For each $x \in \Mbold(e,\rho)$, we have
\begin{eqnarray}
e \xrightarrow{ \Fhat} e' \in \chi^{\Nbold|_{\rho} \Mbold|_{\rho}}_{e}\circ i_{(e,\rho)}(x)
& \iff &
e \xrightarrow{ \Fhat} e' \in \chi^{\Nbold|_{\rho} \Mbold|_{\rho}}_{e}(x)
\nonumber \\
 &\iff &
\Mbold|_{\rho}\left(e \xrightarrow{\Fhat} e'\right)(x) \in \Nbold|_{\rho}(e')
\nonumber \\
 &\iff &
\Mbold\left((e,\rho) \xrightarrow{ \Fhat} (e',\rho)\right)(x) \in \Nbold(e',\rho)
\nonumber \\
 &\iff &
(e,\rho) \xrightarrow{\Fhat} (e',\rho) \in \chi^{\Nbold \Mbold}_{(e,\rho)}(x)
\nonumber \\
 &\iff & 
e \xrightarrow{ \Fhat} e' \in \flat(\chi^{\Nbold \Mbold}_{(e,\rho)}(x))
,
\end{eqnarray}
which implies the commutativity of the diagram (\ref{eq:logicalrelation1}).
In particular, if $\Nbold$ is a projective subobject,
$\chi^{\Nbold \Mbold}_{(e,\rho)}(x)$ is a natural sieve,
i.e., $\chi^{\Nbold \Mbold}_{(e,\rho)}(x)= \natural \chi^{\Nbold \Mbold}_{(e,\rho)}(x)$.
This implies the commutativity of the left half
of the diagram (\ref{eq:logicalrelation2}).
\hfill\qed

Since the map $\flat: \Omegabold(e,\rho) \to \Omegabold_{\rho}(e)$
is order-preserving but in general not injective,
it does not faithfully 
preserve the order structure of 
$\chi^{\Nbold \Mbold}_{(e,\rho)}(\Mbold(e,\rho))$;
the image $\flat(\chi^{\Nbold \Mbold}_{(e,\rho)}(\Mbold(e,\rho)))$,
which equals $\chi^{\Nbold|_{\rho}\, \Mbold|_{\rho}}_{e}(\Mbold|_{\rho}(e))$,
loses some information about the structure of 
$\chi^{\Nbold \Mbold}_{(e,\rho)}(\Mbold(e,\rho))$.
On the other hand,
$\flat': \natural \Omegabold(e,\rho) \to \Omegabold_{\rho}(e)$
is a Heyting-algebra isomorphism.
Therefore, 
the commutativity of the squares
in (\ref{eq:logicalrelation2}) implies that,
if $\Nbold$ is projective for $\Mbold$,
the images $\natural \chi^{\Nbold \Mbold}_{(e,\rho)}(\Mbold(e,\rho))$
($\cong \chi^{\Nbold \Mbold}_{(e,\rho)}(\Mbold(e,\rho))$)
and $\chi^{\Nbold|_{\rho}\, \Mbold|_{\rho}}_{e}(\Mbold|_{\rho}(e))$
possess the same order structure in the same Heyting algebra.
In this sense,
the maps $\natural\chi^{\Nbold \Mbold}_{(e,\rho)}$
(or $\chi^{\Nbold \Mbold}_{(e,\rho)}$ equipped with the
alternative target $\natural \Omegabold(e,\rho)$)
and $\chi^{\Nbold|_{\rho}\, \Mbold|_{\rho}}_{e}$
give logically the same assignments
to any $x \in \Mbold(e,\rho)$
and $x = i_{(e,\rho)}(x) \in \Mbold|_{\rho}(e)$,
respectively. 

Our construction and argumentation 
given in the present section is valid
also for $\Sets^{\Ccal^{(e,\rho)\downarrow }}$
as well as for $\Sets^{\Ccal}$.
In particular, for $\Lbold^{(e,\rho)\downarrow}$
and its true subobject $\Tbold^{e_{r},\rho}$,
we have the objects in $\Sets^{\Ccal_{\rho}}$,
$\Lbold^{(e,\rho)\downarrow}|_{\rho} = \Lbold$
and $\Tbold^{e_{r},\rho}|_{\rho} = \Tbold^{r}_{e}$.
Furthermore, $\Tbold^{e_{r},\rho}$ is projective for 
$\Lbold^{(e,\rho)\downarrow}$,
as is pointed out at the top of this subsection.
We thus have the following commutative diagram
as a special case of ({\ref{eq:logicalrelation2}):
\begin{equation}
\begin{CD}
\Lbold_{(e,\rho)\downarrow}(e,\rho) @ > 
\natural \Vmath^{e_{r},\rho} 
>> \natural\Omegabold_{(e,\rho)\downarrow }(e,\rho) 
  @ > \iota^{\natural\Omegabold_{(e,\rho)\downarrow }\,
    \Omegabold_{(e,\rho)\downarrow }}_{(e,\rho)} >> \Omegabold_{(e,\rho)\downarrow }(e,\rho)\\
   @ V i_{(e,\rho)} VV @ VV \flat' V  @ VV \flat V \\
\Lbold(e)  @ >> \Vcal^{r}_{e} > \Omegabold_{\rho}(e) @ = \Omegabold_{\rho}(e)
\end{CD}
\; \; .
\label{eq:logicalrelation3}
\end{equation}
Here, note that
$\Vcal^{r}_{e} := \chi^{\Tbold^{r}_{e}\, \Lbold}_{e}$,
$\natural \Vmath^{e_{r},\rho} :=
\natural \chi^{\Tbold^{e_{r},\rho} \Lbold_{(e,\rho)\downarrow}}_{(e,\rho)}$,
and
$\Vmath^{e_{r},\rho} := \chi^{\Tbold^{e_{r},\rho} \Lbold_{(e,\rho)\downarrow}}_{(e,\rho)}
= \iota^{\natural\Omegabold_{(e,\rho)\downarrow } \,\Omegabold_{(e,\rho)\downarrow }} 
\circ \natural \Vmath^{e_{r},\rho}$.
As did in the last paragraph,
we conclude that
the valuation functions $\natural \Vmath^{e_{r},\rho}$
(or $\Vmath^{e_{r},\rho}$ equipped with the
alternative target $\natural\Omegabold_{(e,\rho)\downarrow }(e,\rho)
 = \natural \Omegabold(e,\rho)$)
and $\Vmath^{e_{r},\rho}$
assign logically equivalent truth-values
to any quantum proposition $P \in
\Lbold_{(e,\rho)\downarrow}(e,\rho)=\Lbold(e,\rho) = \Lcal$
and the same $P \in \Lbold(e) = \Lcal$,
respectively.


\section{Conclusion}
\label{Conclusion}
We have constructed topos-theoretic truth-value valuations
of quantum propositions
in the functor categories $\Sets^{\CcalR}$ and $\Sets^{\Ccal}$.
They are extension of Bub's modal formulism;
as each true atom $e_{r}$ determines a true subset
of the determinate sublattice $\Dcal(e,R)$
and the corresponding 2-valued valuation $V^{e_{r}}$
defined on $\Dcal(e,R)$,
$e_{r}$ determines a true subobject
and defines $\Vcal^{r}_{e}$ in $\Sets^{\CcalR}$
and $\Vmath^{e_{r}, \rho}$ in $\Sets^{\Ccal}$.

Truth values given by $\Vcal^{r}_{e}$
are sieves of which elements, 
morphisms of the base category $\CcalR$,
are linear operators
commutative with the determinate observable $R$. 
Each quantum proposition $P$ assigned a sieve 
$\Vcal^{r}_{e}(P)$ on $e$ 
consisting of morphisms $\Fhat$ 
such that $\Fhat(P) \geq \Fhat(e_{r})$.

In that morphisms are linear operators, 
our theory is similar to an example 
which Isham~\cite{Isham07} constructed in $\Mbold$-$\Sets$ topos.
In fact, if we modify the base category 
$\CcalR$ by adding the zero-space as an object
and follow the procedure given in Section \ref{sec:Topos-Theoretic Valuations}
to obtain alternative valuations,
then they can be reconstructed by another method
based on $\Mbold$-$\Sets$ generated by $\Com(R)$.

Because of the existence of operators vanishing $e_{r}$,
$\Vcal^{r}_{e}$ does not satisfy 
the null-proposition condition.
We have shown that this defect is removed
if we adopt the subobject semi-classifier 
$\delta_{r} \Omegabold$ instead of $\Omegabold$ 
as the target of $\Vcal^{r}_{e}$.
The subobject semi-classifier 
$\delta_{r} \Omegabold$ is a subobject of $\Omegabold$.
Each $\delta_{r} \Omegabold(e)$ is a Heyting algebra
and a sublattice of $\Omegabold(e)$.
Furthermore, it includes the image of $\Lbold(e)$ by $\Vcal^{r}_{e}$.
The truth value $\bot_{e_{r}}$ of the zero proposition is 
a bottom of $\delta_{r} \Omegabold(e)$.

Also, the notion of subobject semi-classifier 
has been invoked to reconcile the valuations 
in $\Sets^{\Ccal_{\rho}}$'s and $\Sets^{\Ccal}$.
It has been shown that  projective subobjects 
of an object of $\Sets^{\Ccal}$ have a subobject semi-classifier
$\natural \Omegabold$.
Each component $\natural \Omegabold (e,\rho)$, 
which consists of natural sieves,
is a Heyting algebra isomorphic to $\Omega_{\rho}(e)$
and is a sublattice of $\Omegabold(e,\rho)$.
This can be immediately applied to $\Sets^{\Ccal_{(e,\rho)\downarrow }}$
because the true subobject $\Tbold^{e_{r},\rho}$ is projective
for $\Lbold^{(e,\rho)\downarrow }$.
As a result, $\Vmath^{e_{r}, \rho}$ equipped with the subobject semi-classifier
as a target gives equivalent truth values as $\Vcal^{r}_{e}$
for any quantum proposition.

In general,
Heyting-algebra structure of subobject classifiers can be redundant
as a target of truth-value valuations of quantum propositions $\Lcal$.
Therefore it is desirable to reduce the target Heyting algebra 
to smaller ones, provided the logical structure of the image of $\Lcal$
by the valuations is faithfully preserved. 
The subobject semi-classifiers given in sections 3.4 and 5.2
are just the cases.
In particular, the smallest (hence irreducible) one, if exists,
would be regarded as a proper target space of valuation functions.

Finally we note that
we have not addressed any application
to concrete problems;
in the present paper,
we have been concentrated on formulation. 
Bub~\cite{Bub97}, however, applies his formulism to 
various issues concerning the foundations or interpretations
of quantum mechanics.
It would be necessary and significant to examine whether
our topos theoretic formulism is applicable to the issues as well.


\appendix

\section{Subobject Semi-Classifier}
\label{sec:Subobject Semi-Classifier}
Let $\Tcal$ be a topos
with a subobject classifier $\Omega$ and a terminal object $1$.
Suppose that $\Omega$ has a subobject $\Delta \Omega$,
and let $\Delta \Omega \xrightarrow{\iota^{\Delta \Omega \, \Omega}} \Omega \in \Mor(\Tcal)$
be an inclusion morphism, i.e., a monomorphism.
Further suppose that there exists a morphism 
$1 \xrightarrow{\Delta \tau} \Delta \Omega \in \Mor(\Tcal)$
which makes the following diagram a pullback:
\begin{equation}
\begin{CD}
1 @ = 1 \\
   @ V {\Delta \tau} VV @ VV  \tau V \\
 \Delta \Omega @ >>  \iota^{\Delta \Omega \, \Omega} >  \Omega
\end{CD}
\;\; .
\end{equation}
If $\Delta \Omega$ acts as if a subobject classifier
of particular class of subobjects,
we call it a subobject semi-classifier.
Precisely, we give the following definition:
\begin{dfn}
Let $M$ be an object of $\Tcal$
and $\Delta \Sub (M)$ ($\subseteq \Sub (M)$) 
be a collection of its subobjects.
If, for any $N \in  \Delta \Sub (M)$,
there exists a morphism 
$M \xrightarrow{\Delta \chi^{N M}} \Delta \Omega \in \Mor(\Tcal)$
such that
\begin{equation}
\chi^{N M} = \iota^{\Delta \Omega \, \Omega} \circ \Delta \chi^{N M},
\end{equation}
then $\Delta \Omega$ is called a subobject semi-classifier
of $\Delta \Sub (M)$.
\end{dfn}
We can extend this definition to 
a collection of $\Sub (M)$.
\begin{dfn}
Let $\Delta \Obj(\Tcal)$ ($\subseteq \Obj(\Tcal)$) be
a collection of objects such that, 
for any $M \in \Delta \Obj(\Tcal)$,
$\Delta \Omega$ is a subobject semi-classifier of $\Delta \Sub(M)$.
Then $\Delta \Omega$ is called a subobject semi-classifier
of the collection 
$\Delta \Sub := \{\Delta \Sub(M) : M \in \Delta \Obj(\Tcal)\}$.
\end{dfn}

The subobject semi-classifier $\Delta \Omega$ is a via point
where the characteristic morphism $\chi^{NM}$ is factored through.
As is seen in the following propositions, however,
it acts together with $\Delta \tau$ as a subobject classifier 
of $\Delta \Sub(M)$.
The naming is thus justified.
\begin{prp}
Suppose that $M \in \Obj(\Tcal)$ and $N \in \Delta \Sub(M)$,
then the diagram
\begin{equation}
\begin{CD}
N @ >  ! >> 1 \\
   @ V {\iota^{N M}} VV @ VV  \Delta \tau V \\
 M @ >> \Delta \chi^{NM} > \Delta \Omega
\end{CD}
\end{equation}
is a pullback.
\end{prp}
{\bf Proof.}
Consider the following diagram:
\begin{equation}
\begin{CD}
 N @ > ! >> 1 @ =  1     \\
@ V \iota^{NM} VV  @ VV \Delta \tau V  @ VV \tau V \\
M @ >> \Delta \chi^{NM} > \Delta \Omega @ >> \iota^{\Delta \Omega \, \Omega} > \Omega
\end{CD}
\;\; .
\end{equation}
As is previously noted for diagram (A.1), 
the right-half square is a pullback,
and so is the outer square because of equation (A.2).
Therefore, also, the left-half square must be a pullback. 
\hfill\qed

\begin{prp}
(i) 
Suppose that $M \in \Obj(\Tcal)$ and $N \in \Delta \Sub(M)$.
If there exists a morphism 
$M \xrightarrow{\zeta} \Delta \Omega$
which makes the diagram,
\begin{equation}
\begin{CD}
N @ >  ! >> 1 \\
   @ V {\iota^{NM}} VV @ VV \Delta \tau V \\
 M @ >>  \zeta> \Delta \Omega
\end{CD}
\;\;,
\end{equation}
a pullback,
then $\zeta = \Delta \chi^{NM}$.

(ii) Conversely, for each morphism
$M \xrightarrow{\zeta} \Delta \Omega$,
there exist, up to isomorphism, 
subobjects $N \in \Delta \Sub(M)$
satisfying $\zeta = \Delta \chi^{NM}$,
hence, making the corresponding diagram (A.5) a pullback.
\end{prp}
{\bf Proof.}
To prove the statements (i) and (ii),
we use the following diagram:
\begin{equation}
\begin{CD}
 N @ > ! >> 1 @ =  1    \\
@ V \iota^{NM} VV  @ VV \Delta \tau V  @ VV \tau V \\
M @ >> \zeta > \Delta \Omega @ >> \iota^{\Delta \Omega \, \Omega} > \Omega
\end{CD}
\;\; .
\end{equation}

(i) As previously noted, 
the right-half of the diagram (A.6) is a pullback.
Therefore,
if the diagram (A.5),
i.e., the left-half of the diagram (A.6), is a pullback,
then, so is the outer square of (A.6).
But then, because of the uniqueness property of characteristic morphisms in toposes,
$\chi^{NM} = \iota^{\Delta \Omega \, \Omega} \circ \zeta$.
Thus, because of equation (A.2), 
we have 
$\iota^{\Delta \Omega \, \Omega} \circ \zeta 
= \iota^{\Delta \Omega \, \Omega} \circ \Delta \chi^{NM}$.
Since $\iota^{\Delta \Omega \,\Omega}$ is monic, $\zeta = \Delta \chi^{NM}$ follows.

(ii) Since any topos has all finite limits,
we can take a morphism $N \xrightarrow{f} M$ 
up to isomorphism
as a pullback of $1 \xrightarrow{\Delta \tau} \Delta \Omega$ 
along the morphism 
$M \xrightarrow{\zeta} \Delta \Omega$.
Here, since $\zeta$ is monic, so is $N \xrightarrow{f} M$.
Therefore, $N \xrightarrow{f} M$ can be regarded 
as an inclusion morphism, $\iota^{NM}$.
Under these conditions,
consider the diagram (A.6).
Also in this case, the outer square is a pullback, hence, 
$\chi^{NM} = \iota^{\Delta \Omega \, \Omega} \circ \zeta$.
\hfill\qed

%

\section{Propositions on the Functor $\Abold$ in $\Sets^{\Ccal}$}
\label{sec:Propositions on the Functor}

\begin{prp}
Suppose that $\rho \leq \rho'$.
Then, $\Abold(e,\rho) \subseteq \Abold(e,\rho')$ implies 
$\Abold(e,\rho) = \Abold(e,\rho')$
\end{prp}
{\bf Proof.}
The Hilbert space can be decomposed as
$\Hcal = \bigoplus_{i \in I} r_{i}$. 
Here, $\{r_{i}: i \in I \} = \ES(\rho)$,
and, for any $ i \in I $, $ r_{i} = \bigoplus_{j \in J_{i}} r'_{ij} $,
where $\{r'_{ij}: i \in I,\; j \in J_{i}\} = \ES(\rho')$.

If $e_{r'_{ij}} \neq \{0\}$,
then $e_{r_{i}}$ cannot be $\{0\}$ because $r'_{ij} \subseteq r_{i}$.
But then there exists $j' \in J_{i}$ s.t. $e_{r_{i}} = e_{r'_{ij'}}$
because $e_{r_{i}} \in \Abold(e,\rho')$.
If $r'_{ij} \neq  r'_{ij'}$, $e_{r_{i}}$ cannot be $e_{r'_{ij'}}$.
Thus, $r'_{ij} =  r'_{ij'}$, hence, $e_{r'_{ij}} = e_{r_{i}} $.
This implies that $e_{r'_{ij}} \in \Abold(e,\rho)$.

If $e_{r'_{ij}} = \{0\}$, then $e_{r'_{ij}} \in \Abold(e,\rho)$.

Thus, $\Abold(e,\rho') \subseteq  \Abold(e, \rho)$ is proved.
\hfill\qed

\begin{prp}
Suppose that $\Abold(e,\rho) \subseteq \Abold(e,\rho')$ and $\rho \leq \rho'$.
Then for any $\rho''$ satisfying $\rho \leq \rho'' \leq \rho'$,
it follows that
$\Abold(e,\rho) \subseteq \Abold(e,\rho'') \subseteq \Abold(e,\rho')$.
\end{prp}
{\bf Proof.}
The Hilbert space can be decomposed as
$\Hcal = \bigoplus_{i \in I} r_{i}$. 
Here, $\{r_{i}: i \in I \} = \ES(\rho)$,
and, for any $ i \in I $, $ r_{i} = \bigoplus_{j \in J_{i}} r''_{ij} $,
where $\{r''_{ij}: i \in I,\; j \in J_{i}\} = \ES(\rho'')$,
and furthermore,
$ r''_{ij} = \bigoplus_{k \in K_{ij}} r'_{ijk} $, where
$\{r'_{ijk}: i \in I,\; j \in J_{i}, \; k \in K_{ij}\} = \ES(\rho')$.
That is, 
$ r_{i} = \bigoplus_{j \in J_{i}}\bigoplus_{k \in K_{ij}} r'_{ijk} $.

Take any $i \in I$ and fix it.

If $e_{r_{i}} \neq \{ 0 \}$, 
then $\exists \, ! \,j' \in J_{i}$, $\exists \, ! \,k' \in K_{ij'}$
s.t. $e_{r_{i}} = e_{r'_{ij'k'}}$ 
and $e_{r'_{ijk}}= \{0\}$ for any $r'_{ijk}$ except for $r'_{ij'k'}$
because $e_{r_{i}} \in \Abold(e,\rho')$.
But then, for $r''_{ij'} = \bigoplus_{k \in K_{ij'}} r'_{ij'k}$,
$e_{r''_{ij'}}=e_{r'_{ij'k'}}=e_{r_{i}}$,
and for other $r''_{ij}$ $(j \in J_{i})$, $e_{r''_{ij'}}=\{0\}$.
This implies that 
$e_{r_{i}} \in \Abold(e,\rho'')$ and, 
for any $j \in J_{i}$, $e_{r''_{ij}} \in \Abold(e,\rho')$.

If $e_{r_{i}} = \{0\}$, $e_{r''_{ij}} = \{0\}$ for $\forall j \in J_{i}$
and $e_{r'_{ijk}} = \{0\}$ for $\forall j \in J_{i}$ and $\forall k \in K_{ij}$.
Thus also in this case, 
it is shown that $e_{i} \in \Abold(e,\rho'')$ and $e_{r''_{ij}} \in \Abold(e,\rho')$.
\hfill\qed

Note that the inclusion relation $\subseteq $'s 
can in fact be replaced by $=$'s
because of Proposition B1. 


\section{Complement to Proof of Proposition 5.10}
\label{sec:Complement to Proof}
In the proof of Proposition 510,
we use the following proposition:

\begin{prp}
Suppose that $S \in \Omegabold(e,\rho)$ is natural.
Then, for any $\Fhat \in \Com(\rho)$,
\begin{equation}
(e,\rho) \xrightarrow{\Fhat_{\rho \rho'}} (e',\rho') \in S
\; \Longrightarrow \;
(e,\rho) \xrightarrow{\Fhat_{\rho \rho}} (e',\rho) \in S. 
\end{equation}
\end{prp}
{\bf Proof.} 
Since $S$ is natural, i.e., $S = \natural S$,
$\Fhat_{\rho \rho'} \in S$ implies 
that there exist arrows, 
$e \xrightarrow{\Ghat} e'' \in \flat(S)$
and 
$(e'', \rho) \xrightarrow{\Hhat_{\rho \rho'}} (e', \rho') \in \Mor(\Ccal)$,
such that
\begin{equation}
\Fhat_{\rho \rho'} = \Hhat_{\rho \rho'} \circ \Ghat_{\rho \rho}.
\label{eq:FHGappendix}
\end{equation}
On the other hand, we have
\begin{equation}
\Hhat_{\rho \rho'} = \Ihat_{\rho \rho'} \circ \Hhat_{\rho \rho}.
\label{eq:HIHappendix}
\end{equation}

Equations (\ref{eq:FHGappendix}) and (\ref{eq:HIHappendix})
imply that the outer square of the following diagram commutes:
\begin{equation}
\begin{CD}
 (e,\rho) @ > \Ghat_{\rho \rho} >> (e'', \rho) @ > \Hhat_{\rho \rho} >>  (e', \rho)     \\
@ V \Fhat_{\rho \rho'} VV  @ VV \Hhat_{\rho \rho'} V  @ VV \Ihat_{\rho \rho'} V \\
(e',\rho') @ = (e',\rho') @ = (e', \rho')
\end{CD}
\;\;\;.
\end{equation}
Since as operators on $\Hcal$, $\Fhat = \Ihat \Hhat \Ghat = \Hhat \Ghat$,
we have that
\begin{equation}
\Fhat_{\rho \rho} = \Hhat_{\rho \rho} \circ \Ghat_{\rho \rho}.
\label{eq:FHG2appendix}
\end{equation}
Furthermore, since $e \xrightarrow{\Ghat} e'' \in \flat(S)$, 
we have
\begin{equation}
(e,\rho) \xrightarrow{\Ghat_{\rho \rho}} (e'',\rho) \in S.
\label{eq:GinS}
\end{equation}
Equations (\ref{eq:FHG2appendix}) and (\ref{eq:GinS})
imply that $\Fhat_{\rho \rho} \in S$
because $S$ is a sieve.
\hfill\qed




%
%

\end{document}